\documentclass[11pt]{article}

\title{Assessing numerical methods for molecular and particle simulation}

\author{Xiaocheng Shang\footnotemark[1]\ \footnotemark[3] \and Martin Kr\"{o}ger\footnotemark[1] \and Benedict Leimkuhler\footnotemark[2]\ \footnotemark[3] }

\date{\today}

\usepackage{cite}

\usepackage{hyperref}
\usepackage{graphicx}

\usepackage{caption}
\DeclareGraphicsExtensions{.eps,.mps,.pdf,.jpg,.png}
\graphicspath{{figures/}{../figures/}}

\usepackage{fancyhdr}

\usepackage{amsmath}
\usepackage{bm}

\usepackage{amsfonts}

\usepackage{dsfont}

\usepackage[titletoc,toc,title]{appendix}

\usepackage[margin=3.0cm]{geometry}

\newcommand{\E}{\mathbb{E}}

\renewcommand{\vec}[1]{{\mathbf #1}}

\newcommand{\q}{{\vec{q}}}
\newcommand{\p}{{\vec{p}}}
\newcommand{\F}{{\vec{F}}}

\newcommand{\dd}{{\rm d}}

\newcommand{\bea}{\begin{eqnarray}}
\newcommand{\eea}{\end{eqnarray}}
\newcommand{\be}{\begin{equation}}
\newcommand{\ee}{\end{equation}}

\newcommand{\kB}{k_\mathrm{B}}
\newcommand{\sigmaLJ}{r_0}
\newcommand{\sigmaf}{\sigma_\rho}
\newcommand{\mi}{m}

\usepackage{mathrsfs}

\usepackage{soul}
\usepackage{color}

\usepackage{algorithm}
\usepackage{algorithmic}

\usepackage{url}
\usepackage{subfigure}

\usepackage{epstopdf}
\DeclareGraphicsExtensions{.eps,.mps,.pdf,.jpg,.png}

\usepackage{latexsym}
\usepackage{mathrsfs}
\usepackage{amssymb}
\usepackage{enumitem}

\usepackage{url}

\usepackage{array}
\newcolumntype{C}[1]{>{\centering\let\newline\\\arraybackslash\hspace{0pt}}m{#1}}

\begin{document}

\maketitle

\renewcommand{\thefootnote}{\fnsymbol{footnote}}

\footnotetext[1]{Department of Materials, Polymer Physics, ETH Z\"{u}rich, CH-8093 Z\"{u}rich, Switzerland.}
\footnotetext[2]{School of Mathematics and Maxwell Institute for Mathematical Sciences, University of Edinburgh, Edinburgh, EH9 3FD, UK.}
\footnotetext[3]{Corresponding authors. Emails: x.shang@mat.ethz.ch (X. Shang); b.leimkuhler@ed.ac.uk (B. Leimkuhler)}

\begin{abstract}
  We discuss the design of state-of-the-art numerical methods for molecular dynamics, focusing on the demands of soft matter simulation, where the purposes include sampling and dynamics calculations both in and out of equilibrium. We discuss the characteristics of different algorithms, including their essential conservation properties, the convergence of averages, and the accuracy of numerical discretizations. Formulations of the equations of motion which are suited to both equilibrium and nonequilibrium simulation include Langevin dynamics, dissipative particle dynamics (DPD), and the more recently proposed ``pairwise adaptive Langevin'' (PAdL) method, which, like DPD but unlike Langevin dynamics, conserves momentum and better matches the relaxation rate of orientational degrees of freedom. PAdL is easy to code and suitable for a variety of problems in nonequilibrium soft matter modeling; our simulations of polymer melts indicate that this method can also provide dramatic improvements in computational efficiency. Moreover we show that PAdL gives excellent control of the relaxation rate to equilibrium. In the nonequilibrium setting, we further demonstrate that while PAdL allows the recovery of accurate shear viscosities at higher shear rates than are possible using the DPD method at identical timestep, it also outperforms Langevin dynamics in terms of stability and accuracy at higher shear rates.
\end{abstract}

\pagenumbering{arabic}

\section{Introduction}

In this article, we provide a current and detailed perspective on the design of stochastic methods for simulating molecular and particle systems.   Most of our discussion is general and equally applicable to simple and complex molecular fluids and polymer solutions, and to both equilibrium and nonequilibrium modeling.    Modern software packages such as~\mbox{LAMMPS}~\cite{Plimpton1995} offer a bewildering array of options for particle simulation, including choices regarding the model ensemble, equations of motion, discretization method, and parameter selection.   In this article we contrast a number of the different schemes available, drawing on recent advances in the literature and focussing on the practical needs of the simulation community.    All of the existing methods are convergent in the sense that, for suitable choice of parameters and in the limit of small timestep, they are capable of reproducing the exact statistical properties with high accuracy; however, the methods have very different computational efficiencies. In practice, the choice of method can mean the difference between a computational task completing in a day or a week.    The challenge of designing efficient methods is particularly acute in nonequilibrium modeling, where the lack of a simple known form for the invariant distribution makes benchmarking challenging and where the delicate approximation of dynamical behavior plays an important role.

The state-of-the-art in molecular dynamics and its limitations are well documented in recent reviews~\cite{Vlachakis2014,Hassanali2014}.  Let us briefly review the challenges of simulation of polymeric systems~\cite{Gee2006,Kreer2016,Bernardo2016,Li2015c}, which constitute a broad area of research of pharmaceutical, materials, chemical, biological and physical relevance, and an area where simulation times easily exceed available resources. While the system size required to avoid significant finite-size effects \mbox{scales}, with the polymerization degree $N$, as $N^{3/2}$ for flexible chains and as $N^3$ for stiff chains, respectively, the longest relaxation time scales as $N^2$ for dilute systems and as $N^{3}$ (or larger) for concentrated and entangled systems, respectively; the total simulation time is thus $\sim N^6$ for a concentrated polymer solution, while a typical polymerization degree is $N\approx 10^4$--$10^5$ for a synthetic polymer or a biopolymer like hyaluronan. Such systems can be investigated qualitatively using coarse-graining strategies and multiscale modeling approaches of various kinds~\cite{Gooneie2017,Li2013a}. However, atomistic simulation of polymeric systems is limited to the study of a few molecules, or concentrated, eventually semicrystalline systems containing simple polymers like polyethylene with $N\le 2000$ over a duration of a few tens of nanoseconds~\cite{Takahashi2017,Boyd1994,Harmandaris2005,Kavassalis1993,Stephanou2011} and subject to deformation or flow~\cite{Yang2017,Yeh2017,Ramirez-Hernandez2013}. Highly entangled polymeric systems with practically relevant $N$, and their confined counterparts like polymer brushes~\cite{Binder2011,Kreer2016,Posel2017,Speyer2015} are still out of reach for atomistic simulations. The limitations are even more severe for polymers in nanocomposites~\cite{Bernardo2016,Li2014b,Karatrantos2016}, branched or hyperbranched polymers like dendronized \mbox{polymers}~\cite{Le2009,Cordova-Mateo2014,Speyer2015,Bertran2013,Cordova-Mateo2015}, and polyelectrolytes~\cite{Raafatnia2014,Merlitz2015,Das2015}.

This article is addressed to stochastic simulation techniques for polymer models based on generalizations of Brownian or Langevin dynamics. The challenges of simulation are well exemplified by two model polymer systems: (i) the single polymer chain in implicit solvent, and (ii) a polymer melt. Both models combine aspects of sampling and dynamical approximation.   In the case of a single polymer, the dynamics of the thermostat plays a crucial role in describing the relaxation behavior.  It is important in this setting to mimic the underlying internal dynamical processes of the molecule while correctly modeling the exchange of energy between polymer and bath.  For the melt, the key difficult quantities are rheological properties like shear viscosity or normal stress difference and disentanglement time, and the system dynamics is typically dominated by the frequent collisions of particles. We restrict our attention to the polymer melt case in this article, because it is computationally more demanding.

Designing effective algorithms involves the selection of a formulation or modeling framework, choice of parameterization, and design of numerical discretization.  These choices cannot be isolated from each other.  In practice the form of the numerical discretization is strongly dependent on the formulation used, and the choice of parameterization will depend on both the goals of simulation and the numerical scheme.

The common feature of all the most popular formulations in use is that they are designed to facilitate sampling in the canonical ensemble.  That is, when applied to a conservative system with a total number of $N_\mathrm{t}$ particles and Hamiltonian energy function
$H = \sum_{i=1}^{N_\mathrm{t}} \p_{i} \cdot \p_{i}/(2m)  + U(\q_{1},\q_{2}, \dots, \q_{N_\mathrm{t}})$, they are designed to drive the system toward the canonical equilibrium state with probability density
\begin{equation*}
\rho_{\beta}(\q_{1},\q_{2}, \dots, \q_{N_\mathrm{t}}, \p_{1},\p_{2}, \dots, \p_{N_\mathrm{t}}) = Z^{-1} \exp(-\beta H)
\end{equation*}
where $Z$ is a suitable normalizing constant (i.e., the partition function), $\beta=(k_{\rm B}T)^{-1}$, with $k_{\rm B}$ being the Boltzmann constant, $T$ the system temperature,
and $m$ the assumed uniform mass of the $N_\mathrm{t}$ particles.  (Generalizing all results of this article to nonuniform particle masses would be straightforward but the uniform mass simplifies presentation of some formulas.) Temperature control is crucial for $NVT$ simulation, but also plays a fundamental role in barostat methods as these typically fix both temperature and pressure.

A wide range of different approaches can be designed to sample from the canonical distribution; each such formulation has certain desirable properties, but also certain limitations.   For example it is well known that the dynamical response of a system simulated using Langevin dynamics will strongly depend on the friction coefficient.  However, since Langevin dynamics relies on a strong assumption of scale separation, it is possible that there is no choice of the friction that gives a satisfactory dynamical approximation~\cite{Loncharich1992,Feig2007,Bussi2007,Peters2017} (see also discussions on time scales associated with Langevin dynamics~\cite{Peters2017}). In some cases a ``gentle'' form of Langevin dynamics is used where internal relaxation modes within the polymer are the dominant feature of interest, or else one uses Nos\'{e}--Hoover~\cite{Nose1984a,Hoover1985} or stochastic velocity rescaling~\cite{Bussi2007a,Bussi2008,Bussi2009}, both of which are in some sense ``gentle'' alternatives to Langevin dynamics~\cite{Leimkuhler2011}. Other approaches include the pairwise Nos\'e--Hoover thermostat~\cite{Allen2007} and multiparticle collision dynamics~\cite{Winkler2013}. However, we restrict our attention to methods that are derived directly as discretizations of stochastic differential equations.  In case the goal is only to sample the equilibrium structures of the system under study (as often needed in protein modeling~\cite{Maximova2016,Paquet2015}), one may use the coefficient of friction as a free parameter and optimize the choice to enhance the rate of convergence to equilibrium or increase the effective sample size~\cite{Sokal1997} associated with a certain family of observables.
Whereas standard molecular dynamics methods such as Langevin dynamics and its overdamped limit (Brownian dynamics) are appropriate for modeling systems in or near thermodynamic equilibrium, these methods do not take into account the possibility of an underlying flow, and are thus, unless modified, inappropriate for situations where the underlying flow of the system cannot be predicted beforehand (e.g., when dealing with interfaces or nonuniform flow).

More generally, as we coarse-grain the system, the hydrodynamic transport properties  become increasingly important, which calls for a formulation that preserves momentum and Galilean invariance.  A method which addresses these issues is the dissipative particle dynamics (DPD) method of Hoogerbrugge and Koelman~\cite{Hoogerbrugge1992} which conserves momentum ``exactly'' (i.e., up to rounding errors) at each step.   We find in our studies that the recently proposed pairwise adaptive Langevin (PAdL) method~\cite{Leimkuhler2016a}, which mimics DPD in the stationary setting, is preferable to DPD in nonequilibrium applications, e.g., for shear flows.   The use of DPD or PAdL addresses a further problem with Langevin dynamics, namely the nonphysical screening of hydrodynamic interactions observed for Langevin dynamics~\cite{Duenweg1993}.

Once the formulation is chosen, typically a set of stochastic differential equations (\mbox{SDEs}), it is necessary to replace it for computation with a discretized formulation, by applying a specific numerical method.   Particularly for large, computationally demanding applications, the simulation must proceed at the largest possible timestep in order to provide meaningful answers to the questions of interest.   Since we must work within a fixed computational budget, one should ultimately compare different numerical methods in terms of accuracy of observables for fixed computational work.  On the other hand, if, as here, we wish to separate the numerical error (due to discrete approximation) from the statistical error (due to collecting finite numbers of samples) it is better to analyze these two types of error separately.   This is especially true in a metastable system, i.e., one whose dynamics are limited by rare transitions, as for example protein folding or glassy system modeling~\cite{Daggett2002,Barrat2010}, but generally speaking virtually any particle simulation will be subject to the timestep issue when pushed to deliver results on laboratory-relevant timescales.   Thus we are led to seek methods that are optimized to deliver the desired properties in the most cost-effective manner, and the large timestep needed means that discretization bias becomes relevant.   For molecular and mesoscale modeling, the most fundamental type of error incurred is the error in averages.      We say that a method is accurate for long term averaging if the distribution generated by the numerical method converges in the limit of large time and small timestep to a stationary distribution which is close, in the sense of distributions, to the corresponding stationary distribution of the SDE~\cite{Leimkuhler2013,Leimkuhler2013c}.  The other major source of error in large scale simulations can be viewed as the error due to insufficient sampling of a stochastic quantity.   Such error is always present but may be difficult to estimate, since in practice we do not necessarily know the underpinning distribution a priori nor the normalization constant needed to compute a robust average.   To put this another way, if we remain for our entire simulation time exploring certain states of the molecule, it might be the consequence of a strong local confinement rather than an indication that these are always the most important group of states.    Analyzing and comparing numerical methods for sampling thus requires balancing the issues of accuracy and sampling efficiency and it is crucial to realize that different methods may have very different levels of bias and rates of convergence which are highly dependent on the choice of timestep.  In this article we explore these issues in detail for model systems, calculating first the bias and using this to select timestep to control the attainable accuracy, then (for the specific choice of timestep) assessing the rate of convergence to equilibrium average.

The key findings of this article are as follows.  First, we compare the performance of Langevin dynamics, DPD, and PAdL for several benchmark calculations, in particular showing that the PAdL method provides higher accuracy (for given computational budget) than Langevin dynamics and DPD, in both equilibrium and nonequilibrium applications, with careful study of the trade-off between numerical bias and convergence rate in simulation.     Second, we compare our numerical schemes with exact results regarding the relaxation behavior for a benchmark model, thus clarifying the performance of the methods in dynamics-oriented modeling applications. Third, we develop a careful procedure to quantify the sampling efficiency of various methods by comparing the effective sample size. A fourth advance in the current article is the demonstration that PAdL allows the recovery of accurate shear viscosity using larger shear rates than otherwise are possible using DPD (at identical timestep) while PAdL outperforms Langevin dynamics in terms of stability and accuracy at higher shear rates. Finally, we emphasize that this article provides specific details regarding implementation of all the various methods which are often lacking in the literature.

The rest of the article is organized as follows.  In Section~\ref{sec:Numerical_Methods}, we review a variety of numerical methods in polymer melts simulation.  We describe, in Section~\ref{sec:Quantifiers_Performance}, various physical quantities that are used to evaluate the simulation performance of each method. Section~\ref{sec:Numerical_Experiments} presents  numerical experiments in both equilibrium and nonequilibrium cases, comparing the performance of numerous popular numerical methods in practical examples. Our findings are summarized in Section~\ref{sec:Conclusions}.

\section{Numerical methods}
\label{sec:Numerical_Methods}

In this section, we describe various numerical methods used to simulate  many-particle systems.    We are interested both in the choice
of formulation of the equations of motion and in the consequent secondary choice of discretization method.     In the literature one observes that virtually all the popular methods are of a relatively simple design and require typically a single evaluation of the forces of interaction at each timestep, the computational cost of the force evaluation being normally the unit of computational effort.

\subsection{Langevin thermostat}

Following the seminal work of Grest and Kremer~\cite{Grest1986,Kremer1990}, Langevin dynamics has been widely used in simulating Lennard-Jones systems including polymer chains and their melts and can be written as
\begin{equation}
  \label{eq:Langevin}
  \begin{aligned}
    \dd \q_{i} & =  \mi^{-1} \p_{i} \dd t \, , \\
    \dd \p_{i} & =  \F_{i}(\q,t) \dd t - \gamma \left( \mathbf{p}_{i}  - \mi\mathbf{u}_i \right)  \dd t + \sigma \mi^{1/2}\dd {\bf W}_{i} \, ,
  \end{aligned}
\end{equation}
where $\q_{i}$ and $\p_{i}$ are $d$-dimensional vectors and respectively represent positions and absolute momenta of bead $i$ with $d$ being the underlying dimensionality of the physical space (typically $d=3$), $\mi$ denotes the mass of a particle, the force on particle $i$, $\F_{i}(\q,t)$, could in principle be both positions and time dependent, however, in equilibrium, $\F_{i} = -\nabla_{\q_{i}} U$ is the conservative force given in terms of a potential energy function $U=U(\q)$, $\dd{\bf W}_{i}$ represents a dimensionless vector of $d$ independent increments of Wiener processes with stochastic properties $\langle\dd{\bf W}_{i}(t)\rangle={\bf 0}$ and $\langle \dd{\bf W}_{i}(t)\dd{\bf W}_{j}(t')\rangle = \delta_{ij} \delta(t-t'){\bf 1}\dd t$, $\gamma$ is the bead friction coefficient, which couples the beads weakly to a heat bath, ${\bf u}_i={\bf u}({\bf q}_i)$ denotes a macroscopic streaming velocity at position ${\bf q}_i$,
and $\sigma$ represents the strength of the random forces, satisfying the following fluctuation-dissipation relation:
\begin{equation}\label{eq:FDT}
  \sigma^2=2\gamma k_{\mathrm{B}}T \, .
\end{equation}
It should be emphasized here that the damping term in~\eqref{eq:Langevin} depends on the peculiar velocity, which is the difference between the absolute velocity $\mathbf{v}_{i}={\bf p}_i/\mi$ and the streaming velocity field $\mathbf{u}$. That is, the thermostat acts only on the peculiar velocity, which is essential in nonequilibrium, for instance, when modeling shear flow. The traditional formulation of Langevin dynamics (i.e., when $\mathbf{u}={\bf 0}$) does not take into account the underlying streaming velocity and thus is expected to fail when there exists a nonzero underlying streaming velocity, however, this feature is not always clearly stated~\cite{Soddemann2003}.

\subsubsection{Stochastic velocity Verlet (SVV)}

Due to its ease of implementation and its natural construction based on the popular Verlet method of molecular dynamics, the stochastic velocity Verlet (SVV) method~\cite{Melchionna2007} is one of the most popular methods for Langevin dynamics.  The equations are:
\begin{align*}
    \p^{n+1/2}_{i} &= \p^{n}_{i} - \frac{h}{2} \nabla_{\q_{i}} U(\q^{n}) - \frac{h}{2} \gamma \left( \mathbf{p}^{n}_{i}  - \mi \mathbf{u}_i \right) + \sqrt{\frac{h\mi}{2}} \sigma \mathbf{R}^{n}_{i} \, ,
    \nonumber\\
    \q^{n+1}_{i} &= \q^{n}_{i} + h \mi^{-1} \mathbf{p}^{n+1/2}_{i} \, , \\
    \p^{n+1}_{i} &= \p^{n+1/2}_{i} \!\!- \frac{h}{2} \nabla_{\q_{i}} U(\q^{n+1}) - \frac{h}{2} \gamma \left(\!\! \mathbf{p}^{n+1/2}_{i}\!\!-\!\mi \mathbf{u}_i \!\!\right)\! +\! \sqrt{\frac{h\mi}{2}} \sigma \mathbf{R}^{n+1/2}_{i},\nonumber
\end{align*}
where $h$ is the integration timestep, $\mathbf{R}^{n}_{i}$ and $\mathbf{R}^{n+1/2}_{i}$ are vectors of uncorrelated Gaussian white noise with zero mean and unit variance, resampled at each step.

\subsubsection{The BAOAB method}

Numerical integration methods for Langevin dynamics have been studied systematically in terms of the long term sampling performance in recent works of Leimkuhler and Matthews~\cite{Leimkuhler2013,Leimkuhler2013a}.  Of note is the observation that a particular choice of splitting method, ``BAOAB'', based on a Trotter factorization of the stochastic vector field of the system into exactly solvable subsystems, is far superior to  alternative methods in terms of sampling configurational quantities.
The BAOAB method relies on separating the vector field of the system:
\begin{equation}\label{eq:Splitting_LD}
  \dd \left[ \begin{array}{c} \q_{i} \\ \p_{i} \end{array} \right] =  \underbrace{\left[ \begin{array}{c} \mi^{-1}\p_{i} \\ \vec{0} \end{array} \right] \dd t}_\mathrm{A} + \underbrace{\left[ \begin{array}{c} \vec{0} \\  -\nabla_{\q_{i}} U \end{array} \right] \dd t }_\mathrm{B} + \underbrace{\left[ \begin{array}{c} \vec{0} \\ - \gamma \left( \mathbf{p}_{i} - \mi \mathbf{u}_i \right) \dd t + \sigma \mi^{1/2} \dd\mathrm{{\bf W}}_{i} \end{array} \right] }_\mathrm{O} \, ,
\end{equation}
in such a way that each piece can be solved ``exactly''. It is straightforward to solve the ``A'' and ``B'' pieces, respectively. Moreover, it is possible to derive the exact solution to the Ornstein--Uhlenbeck (``O'') part,
\begin{equation}\label{eq:OU}
  \dd \p_{i} = \gamma \mi \mathbf{u}_i \dd t - \gamma \p_{i}\dd t + \sigma \mi^{1/2} \dd\mathrm{{\bf W}}_{i} \, ,
\end{equation}
as
\begin{equation}\label{eq:OU_Exact_Sol}
  \p_{i}(t) = \mi \mathbf{u}_i +  \left( \p_{i}(0) - \mi \mathbf{u}_i \right) e^{-\gamma t} + \sigma \sqrt{ \frac{m\left(1-e^{-2\gamma t}\right)}{2\gamma} }\mathrm{\bf R}_{i} \, ,
\end{equation}
where $\mathrm{\bf R}_{i}$ is a vector of independent and identically distributed (i.i.d.) standard normal random variables. The BAOAB method then can be defined as
\begin{equation}
  e^{h \hat{\mathcal{L}}_\mathrm{BAOAB} } = e^{ (h/2) \mathcal{L}_\mathrm{B} } e^{ (h/2) \mathcal{L}_\mathrm{A} } e^{ h \mathcal{L}_\mathrm{O} } e^{ (h/2) \mathcal{L}_\mathrm{A} } e^{ (h/2) \mathcal{L}_\mathrm{B} }\, ,
\end{equation}
where $\exp\left(h \mathcal{L}_f\right)$ denotes the phase space propagator associated with the corresponding vector field $f$. The integration steps of the BAOAB method, modified to include the streaming velocity, reads:
\begin{align*}
    \p^{n+1/2}_{i} &= \p^{n}_{i} - (h/2) \nabla_{\q_{i}} U(\q^{n}) \, , \nonumber\\
    \q^{n+1/2}_{i} &= \q^{n}_{i} + (h/2) m^{-1}{\p}^{n+1/2}_{i} \, , \nonumber\\
    \tilde{\p}^{n+1/2}_{i} &= \mi \mathbf{u}_i +  \left( \p^{n+1/2}_{i} - \mi \mathbf{u}_i \right)e^{-\gamma h} + \sqrt{\mi  k_{\mathrm{B}}T (1-e^{-2\gamma h})} \, \mathbf{R}^{n}_{i} \, , \\
    \q^{n+1}_{i} &= \q^{n+1/2}_{i} + (h/2)m^{-1}\tilde{\p}^{n+1/2}_{i} \, , \nonumber\\
    \p^{n+1}_{i} &=\tilde{\p}^{n+1/2}_{i} - (h/2) \nabla_{\q_{i}} U(\q^{n+1}) \, . \nonumber
\end{align*}
It should be noted that only one force calculation is required at each step for BAOAB (i.e., the force computed at the end of each step will be reused at the beginning of the next step), the same as for alternative schemes, including the SVV method.

\subsection{Dissipative particle dynamics (DPD) thermostat}

Momentum conservation is an essential property required to correctly capture hydrodynamic interactions. However, the momentum is not conserved in Langevin dynamics due to the fact that the thermostat (i.e., the dissipative and random forces) is not pairwise. Analogous to Langevin dynamics, the dissipative particle dynamics (DPD) method~\cite{Hoogerbrugge1992,Espanol1995,Groot1997} is a momentum-conserving thermostat which has been proposed to simulate complex hydrodynamic behavior. Unlike Langevin dynamics, the dissipative force in DPD is dependent of relative velocities and both the dissipative and random forces are pairwise, ensuring the momentum conservation. It should be noted that DPD has been used primarily as a mesoscale coarse-graining technique, where each DPD particle represents a blob of molecules, however, it has also been used in simulating polymer melts, in which case each DPD particle corresponds to one bead (e.g., see Refs.~\citen{Pastorino2007,Cao2012,Fedosov2010}).

The equations of motion of the DPD system can be written as
\begin{equation*}
  \label{eq:DPD}
  \begin{aligned}
    \dd \mathbf{q}_{i} &= \mi ^{-1}\mathbf{p}_{i} \dd t \, , \\
    \dd \mathbf{p}_{i} &= \sum_{j\neq i}\left[\mathbf{F}_{ij}^{\mathrm{C}}(r_{ij}) \dd t - \gamma \omega^{\mathrm{D}}(r_{ij})(\mathbf{e}_{ij}\cdot \mathbf{v}_{ij})\mathbf{e}_{ij} \dd t
     + \sigma \omega^{\mathrm{R}}(r_{ij})\mathbf{e}_{ij} \dd \mathrm{W}_{ij}\right] \, ,
  \end{aligned}
\end{equation*}
where $r_{ij} = \| \mathbf{q}_i - \mathbf{q}_j\|$ is the distance between particles $i$ and $j$ with $\mathbf{e}_{ij} = (\mathbf{q}_{i} - \mathbf{q}_{j})/r_{ij}$ being the unit vector in the associated direction, ${\bf v}_{ij}={\bf v}_i-{\bf v}_j$ is the relative velocity, $\mathbf{F}_{ij}^{\mathrm{C}}(r_{ij})$ denotes the conservative force derived from the corresponding pair potential energy $U(r_{ij})$, and $\dd \mathrm{W}_{ij}=\dd \mathrm{W}_{ji}$ are independent increments of Wiener processes with mean zero and variance $\dd t$.  In addition to the relation in~\eqref{eq:FDT}, the two weight functions have to be related by $\omega^{\mathrm{D}}(r_{ij})=\left[\omega^{\mathrm{R}}(r_{ij})\right]^{2}$ in order for the system to sample the canonical ensemble.

We have observed~\cite{Leimkuhler2015} that standard DPD methods perform similarly in all the quantities that we have tested. Therefore, following Ref~\citen{Leimkuhler2016a}, Shardlow's S1 splitting method (i.e., the DPD-S1 scheme)~\cite{Shardlow2003} was used to represent the standard DPD formulation. As in Langevin dynamics, we can similarly define the DPD-S1 (OBAB) method as
\begin{equation}
  e^{h \hat{\mathcal{L}}_\mathrm{DPD-S1} } = e^{ h \mathcal{L}_\mathrm{O} } e^{ (h/2) \mathcal{L}_\mathrm{B} } e^{ h \mathcal{L}_\mathrm{A} } e^{ (h/2) \mathcal{L}_\mathrm{B} }\, ,
\end{equation}
where one should note that the ``O'' part is further split into interacting pairs and then each pair is solved by using the method of Br\"{u}nger, Brooks, and Karplus (BBK)~\cite{Brunger1984} (the detailed integration steps of the DPD-S1 scheme can be found in Appendix~\ref{sec:Appendix_Schemes}).

Due to the fact that the dissipative force depends on relative velocities, DPD is Galilean-invariant, which makes it a profile-unbiased thermostat (PUT)~\cite{Evans1986,Evans2008} by construction and an ideal thermostat for nonequilibrium molecular dynamics (NEMD)~\cite{Soddemann2003}. The PUT allows the simulation itself to define the local streaming velocity (for more details, see Refs.~\citen{Evans1986,Evans2008,Whittle2010}) and thus there is no need to additionally subtract the underlying streaming velocity in nonequilibrium applications.

\subsection{Pairwise adaptive Langevin (PAdL) thermostat}

Inspired by recent developments in adaptive thermostats~\cite{Jones2011,Leimkuhler2015a,Shang2015}, the pairwise adaptive Langevin (PAdL) thermostat, which can be viewed as ``adaptive DPD'', has been proposed by Leimkuhler and Shang~\cite{Leimkuhler2016a}, see also more discussions therein. It has been observed that PAdL is able to correct for thermodynamic observables while mimicking the dynamical properties of DPD.

The equations of motion of the momentum-conserving PAdL thermostat are given by
\begin{equation*}
  \label{eq:PAdL}
  \begin{aligned}
    \dd \mathbf{q}_{i} &= \mi^{-1}\mathbf{p}_{i} \dd t \, , \\
    \dd \mathbf{p}_{i} &= \sum_{j\neq i}\left[\mathbf{F}_{ij}^{\mathrm{C}}(r_{ij}) \dd t - \xi \omega^{\mathrm{D}}(r_{ij})(\mathbf{e}_{ij}\cdot \mathbf{v}_{ij})\mathbf{e}_{ij} \dd t + \sigma \omega^{\mathrm{R}}(r_{ij})\mathbf{e}_{ij} \dd \mathrm{W}_{ij}\right] \, , \\
    \dd \xi            &= G(\mathbf{q},\mathbf{p})\dd t \, ,
  \end{aligned}
\end{equation*}
where $\xi$ is an auxiliary dynamical friction variable,  $\sigma$ a constant amplitude as in Langevin dynamics, and $G(\mathbf{q},\mathbf{p})$ denotes the accumulated deviation of the instantaneous temperature away from the target temperature
\begin{equation}
  \label{eq:PNHL_G}
    G(\mathbf{q},\mathbf{p}) = \frac{1}{\mu} \sum_{i}\sum_{j>i}\omega^{\mathrm{D}}(r_{ij}) \left[ \left( \mathbf{v}_{ij}\cdot \mathbf{e}_{ij} \right)^{2} - 2k_{\mathrm{B}}T/m \right] \, ,
\end{equation}
where $\mu$ is a coupling parameter (an inverse surface mass density) which is referred to as the ``thermal mass''. It can be shown that, in equilibrium, the PAdL system preserves the momentum-constrained canonical ensemble with a modified density
\begin{equation}\label{eq:Gibbs_PAdL}
  \begin{aligned}
  \tilde{\rho}_{\beta}(\mathbf{q},\mathbf{p},\xi) & = \,  \frac{1}{Z} \exp\left[{-\beta H(\mathbf{q},\mathbf{p})}{-\frac{\beta\mu}{2}(\xi-\gamma)^{2}}\right] \\
  & \,  \times \delta\!\left( \sum_i p^{x}_{i}- \pi_x \right) \delta\!\left (\sum_i p^{y}_{i} - \pi_y \right) \delta\!\left (\sum_i p^{z}_{i} - \pi_z \right)  \, ,
  \end{aligned}
\end{equation}
where $\gamma$ is the friction coefficient as it satisfies the
fluctuation-dissipation relation~\eqref{eq:FDT}, and $\mathbf{\pi}=(\pi_x,\pi_y,\pi_z)$ is the linear momentum vector. Additional modifications should be included if the angular momentum is also conserved.

According to the invariant distribution~\eqref{eq:Gibbs_PAdL}, the auxiliary variable $\xi$ is Gaussian distributed with mean $\gamma$ and variance $(\beta \mu)^{-1}$. That is, the auxiliary variable will fluctuate around its mean value during simulation and moreover we can vary the value of the friction in order to recover the dynamics of DPD in a wide range of friction regimes. Therefore, the PAdL thermostat can be viewed as the standard DPD system with an adaptive friction coefficient (i.e., an adaptive DPD thermostat). Furthermore, we point that the PAdL thermostat inherits key properties of DPD (such as Galilean invariance and momentum conservation) required for consistent hydrodynamics. Note also that the PAdL thermostat would effectively reduce to the standard DPD formulation in the large thermal mass limit (i.e., $\mu \rightarrow \infty$).

The splitting method of PAdL proposed in Ref.~\citen{Leimkuhler2016a} has been adopted in this article:
\begin{equation*}
  e^{h \hat{\mathcal{L}}_\mathrm{PAdL} } = e^{ (h/2) \mathcal{L}_\mathrm{A} } e^{ (h/2) \mathcal{L}_\mathrm{B} } e^{ (h/2) \mathcal{L}_\mathrm{O} } e^{ h \mathcal{L}_\mathrm{D} } e^{ (h/2) \mathcal{L}_\mathrm{O} } e^{ (h/2) \mathcal{L}_\mathrm{B} } e^{ (h/2) \mathcal{L}_\mathrm{A} }\, ,
\end{equation*}
where the ``O'' part is again further split into interacting pairs but each pair is solved exactly, and ``D'' represents the additional Nos\'{e}--Hoover part (the detailed integration steps of the PAdL method can also be found in Appendix~\ref{sec:Appendix_Schemes}). It is worth mentioning that the computational costs per timestep of all the methods examined in this article are very similar. With the help of computation-saving devices such as Verlet neighbor lists~\cite{Verlet1967}, the computational effort of all methods under study scales with the number of particles, and both DPD and PAdL are only slightly more expensive than Langevin integrators.

\section{Quantifiers for simulation performance}
\label{sec:Quantifiers_Performance}

In this section, we briefly outline the quantities we will compute in simulation and use to compare the performance of different simulation schemes.  These divide into observables for  equilibrium and nonequilibrium sampling and the rate of convergence to equilibrium.

\subsection{Summed autocorrelation count (SAC)}
\label{subsubsec:SAC}

As in Markov chain Monte Carlo methods, we are interested in accurately and efficiently estimating the expected value of some physical observable of interest $f(x)$, i.e.,
\begin{equation}
  \E[f] = \langle f \rangle = \int_{\Omega_{x}} f(x) \rho(x) \, \dd x \, ,
\end{equation}
by averaging a time series of (typically correlated) $N_\mathrm{s}$ samples
\begin{equation}\label{eq:sample_mean}
  \overline{f} = \frac{1}{N_\mathrm{s}} \sum^{N_\mathrm{s}}_{i=1} f(x_{i})
\end{equation}
for large $N_\mathrm{s}$, where $x_i$ denotes the phase space configuration at time $t_i$. Denoting the variance of $f$ with respect to the probability density function $\rho$ by $\sigmaf^2$, which is independent of particular sampling methods, it can be shown~\cite{Berg2004} that the variance of the estimator $\overline{f}$ of the mean is
\begin{equation}
  \overline{\sigma}^{2}_\rho\left(\overline{f}\right) = \frac{\tau\sigmaf^{2}}{N_\mathrm{s}} \, ,
\end{equation}
where $\tau$ indicates a quantity that we refer to as the ``summed autocorrelation count'' (SAC)\footnote{Note that the SAC is often referred to as the ``integrated autocorrelation time''~\cite{Sokal1997,Goodman1989,Berg2004} in the computational statistics literature. The problem with using such a term here relates to the fact that the time is a well-defined physical quantity whereas the formula quantifies a number of steps of an iterative procedure.
Since we are mostly interested in how quickly the samples decorrelate in terms of the number of steps, we use the word ``count'' in order to avoid confusion with ``physical time''.}, for correlated samples, typically estimated as $\tau\equiv\tau(k_\mathrm{max})$ for some finite $k_\mathrm{max}<N_\mathrm{s}$ by the ``running'' autocorrelation count estimator\footnote{It should be noted that MATLAB's ``autocorr'' function, despite its name, does not calculate the autocorrelation function \eqref{eq:AF}, however, $1+\tau(k_\mathrm{max})$ is just two times the first $1+k_\mathrm{max}$ ``autocorr'' lags in MATLAB.}
\begin{equation}\label{eq:SAC}
  \tau(k) =  1 + 2 \sum^{k}_{j=1} \left( 1 - \frac{j}{N_\mathrm{s}}\right) \frac{C(j)}{C(0)} \, ,
\end{equation}
where
\begin{equation}\label{eq:AF}
  C(j) = \frac{1}{N_\mathrm{s}-j} \sum^{N_\mathrm{s}-j}_{i=1} \left( f(x_{i}) - \overline{f} \right) \left( f(x_{i+j}) - \overline{f} \right),
\end{equation}
is the unnormalized autocorrelation function (or auto co-variance) of $f$. If the samples are uncorrelated, $\tau=1$, in which case the variance of the estimator $\overline{f}$ would simply be $\sigmaf^{2}/N_\mathrm{s}$.
Note that the variance of $f$ (i.e., $\sigmaf^{2}$) is a special case of the autocorrelation~\eqref{eq:AF}, i.e., $\sigmaf^{2} = C(0)$.
The running $\tau(k)$ starts at unity for $k=0$,
vanishes exactly for $k=N_\mathrm{s}-1$, and its value and variance go through a maximum for intermediate $k$. In our simulations, we found that it was unclear how to properly determine the $k_\mathrm{max}$. Therefore, we instead suggest to approximate the SAC based on a weighted sum fitting~\cite{Aust2002} of the normalized autocorrelation function, whose argument is scaled to physical time, $t$, for convenience:
\begin{equation}\label{eq:Fitting_Weighted_Sum}
  \frac{C(t)}{C(0)} = (1-c)e^{-t/\lambda_1} + c  \left(\cos(wt)+\frac{\sin(wt)}{w\lambda_2}\right)e^{-t/\lambda_2} \, ,
\end{equation}
where $c\in[0,1]$ is dimensionless, $\lambda_1,\lambda_2>0$ are time constants, and $w$ is a frequency. Integrating~\eqref{eq:Fitting_Weighted_Sum} from zero to infinity yields
\begin{equation}\label{eq:Fitting_Integral}
  I \equiv \int^{+\infty}_{0} \frac{C(t)}{C(0)} \, \dd t = (1-c)\lambda_1 + \frac{2c\lambda_2}{1+(\lambda_2w)^2} \, ,
\end{equation}
based on which the SAC can be approximated as
\begin{equation}\label{eq:Fitting_SAC}
  \tau \approx \frac{2I}{h}-1 \, ,
\end{equation}
where $h$ is the stepsize associated with the numerical method. The choice of the functional form~\eqref{eq:Fitting_Weighted_Sum} is somewhat arbitrary, as it simply corresponds to the $c$-weighted superposition of solutions of a damped harmonic oscillator and a monoexponential relaxation process, but we observed good agreement with the actual autocorrelation function behavior in our simulation experiments (see the next section).

It should be noted that the SAC is closely related to the statistical error bar, i.e., the statistical error bar of the estimated mean is the standard deviation of the time series divided by the ``effective sample size'' defined as $N_\mathrm{s}/\tau$. Thus, the SAC is an estimate of the number of iterations, on average, for an independent sample to be drawn, given a correlated chain. Therefore, the SAC directly measures the efficiency of the sampling---a lower value of SAC corresponds to a larger effective sample size, i.e., a more efficient sampling.

\subsection{Configurational temperature}

Since we are mostly interested in configurational sampling, in calculating the SAC we choose for $f$ the configurational temperature~\cite{Rugh1997,Butler1998,Braga2005,Allen2006,Travis2008},
\begin{equation}\label{eq:Config_Temp_f}
  k_{\mathrm{B}} f_{T} = \frac{  \left\langle \nabla U(\q) \cdot \nabla U(\q) \right\rangle }{ \left\langle \nabla^{2}U(\q) \right\rangle }\, ,
\end{equation}
an observable function solely depending on positions whose average in the canonical ensemble, as the kinetic temperature, is precisely the target temperature:
\begin{equation}\label{eq:Config_Temp}
  T = \left\langle f_{T} \right\rangle \, ,
\end{equation}
where $\nabla U$ and $\nabla^{2}U$ respectively denote the gradient and Laplacian of the potential energy $U$ in the configurational phase space (further discussions in Ref.~\citen{Leimkuhler2015}). The corresponding unnormalized correlation function and running SAC are denoted by $C_T(k)$ and $\tau_T(k)$, respectively.

\subsection{Polymer conformation}
\label{subsubsec:Correlations}

A multibead nonlinear spring model was employed to simulate a polymer melt as described in detail in Section \ref{subsec:Simulation_Details}.
Denote the coordinate of the $j$-th bead in chain $\alpha$ as $\q^{(\alpha)}_{j}$. The end-to-end vector of chain $\alpha$ is then given by
${\bf R}_\mathrm{ee}^{(\alpha)}={\bf q}_N^{(\alpha)} -{\bf q}_1^{(\alpha)}$. It should be emphasized here that in taking differences one has to respect the periodic boundary conditions (or to simply unfold all polymer contours before applying the above definitions if they are not already kept unfolded within the code).


Correlation functions, which characterize the relevant dynamical properties, are often \mbox{studied} in molecular dynamics. Of particular interest in polymer melts is the orientational autocorrelation function (OAF) of the end-to-end vector of polymer chains, which characterizes the relaxation of the polymer chains and is evaluated by choosing for $f$ the end-to-end vector $\mathbf{R}^{(\alpha)}_{\mathrm{ee}}(t)$ of chain $\alpha$, while all $M$ chains contribute to $C_\mathrm{ee}(t)$ as individual samples. For this vector-valued $f$ the product in Eq.\ (\ref{eq:AF}) is a scalar product. Due to head-tail symmetry $\langle{\bf R}_\mathrm {ee}^{(\alpha)}\rangle={\bf 0}$ and thus $\overline{f}=0$ in that case.

\subsection{Shear viscosity}
\label{subsubsec:Shear_Viscosity}

A common approach to generate a simple shear flow in nonequilibrium molecular dynamics is to apply the well-known Lees--Edwards boundary conditions (LEBC)~\cite{Lees1972}, where, as in normal periodic boundary conditions (PBC), the primary cubic box remains centered at the origin, however, a uniform shear velocity profile is expected~\cite{Evans2008}
\begin{equation}\label{eq:LEBC_Streaming_V}
  \mathbf{u}_i = \dot{\gamma}({\bf q}_i\cdot{\bf e}^y)\mathbf{e}^{x} = \boldsymbol{\kappa}\cdot{\bf q}_i,
  \qquad \boldsymbol{\kappa} = \dot{\gamma}\,{\bf e}^x\otimes{\bf e}^y
\end{equation}
where $\mathbf{e}^{x}$ and $\mathbf{e}^{y}$ respectively denote the unit vector in the $x$- and $y$-direction, $\boldsymbol{\kappa}$ is the transposed velocity gradient tensor, $\otimes$ represents the dyadic product of two vectors, and $\dot{\gamma}$ is the shear rate defined as $\dot{\gamma} = \dd u^{x}/\dd y$, where $u^{x}$ is the macroscopic velocity in the $x$-direction. It is worth mentioning that while LEBC is typically applied only in the $x$-direction, the other directions ($y$ and $z$) remain with PBC. It is nontrivial to implement LEBC in pairwise thermostats due to the position-dependence on both dissipative and random forces, this issue has been  discussed in Ref.~\citen{Leimkuhler2016a}.

The Irving--Kirkwood stress tensor~\cite{Irving1950} subject to LEBC can be written as
\begin{equation}\label{eq:Stree_Tensor}
    \boldsymbol{\sigma} = - \frac{1}{V} \left( \sum_{i} \mi  \left( \mathbf{v}_{i}-\mathbf{u}_i \right) \otimes \left( \mathbf{v}_{i}-\mathbf{u}_i \right) + \sum_{i} \sum_{j>i} \mathbf{q}_{ij} \otimes \mathbf{F}_{ij} \right) \, ,
\end{equation}
where $V$ is the volume of the simulation box, and $\mathbf{u}_{i}$ is the streaming velocity~\eqref{eq:LEBC_Streaming_V} corresponding to the location of particle $i$. Only the conservative force should be included for $\mathbf{F}_{ij}$ in Langevin dynamics since both the dissipative and random forces are averaged out, whereas all three components of the force should be accounted for pairwise thermostats. The generally non-Newtonian shear viscosity is extracted at finite rates as
\begin{equation}\label{eq:Shear_Viscosity}
    \eta = \frac{\langle \sigma_{xy} \rangle}{\dot{\gamma}} \, ,
\end{equation}
where $\sigma_{xy}$ denotes the shear stress, which is the off-diagonal $xy$-component of the symmetric stress tensor $\boldsymbol{\sigma}$~\eqref{eq:Stree_Tensor}. While employing \eqref{eq:Shear_Viscosity} the zero shear viscosity $\eta_0=\lim_{\dot{\gamma}\rightarrow 0}\eta$ can be obtained by extrapolation, it is worth mentioning that $\eta_0$ can be alternatively calculated by integrating the stress-stress autocorrelation function (i.e., the Green--Kubo formulas~\cite{Green1954,Kubo1957}). However, it is well documented that those equilibrium approaches are subject to significant statistical error and thus not preferred in practice (see a detailed discussion on extracting transport coefficients by various approaches in Ref.~\citen{Shang2015a}).

\subsection{Flow alignment angle}
\label{subsubsec:Alignment_Angle}

As a nontrivial application to demonstrate the performance of sampling schemes in the nonequilibrium context, we study the flow alignment of the polymer segments to the imposed flow as a function of the shear rate imposed using LEBC in manner described in Section~\ref{subsubsec:Shear_Viscosity}, thus we calculate the flow alignment angle~\cite{Kroeger1993} as follows:
\begin{equation}\label{eq:Alignment_Angle}
  \chi = \frac{\pi}{4} - \frac{1}{2}\arctan\left( \frac{ \langle b^{2}_{x} - b^{2}_{y} \rangle }{ 2\langle b_{x}b_{y}\rangle } \right),
\end{equation}
where $b_{x}$ and $b_{y}$ respectively represent the $x$- and $y$-component of a normalized bond vector $\mathbf{b} = b_{x} {\bf e}^x + b_{y} {\bf e}^y + b_{z} {\bf e}^z$, and the average is taken over all bonds.

\section{Numerical experiments}
\label{sec:Numerical_Experiments}

In this section, we conduct systematic numerical experiments to compare the performance of various methods introduced in Section~\ref{sec:Numerical_Methods} in polymer melts simulations.

\subsection{Simulation details}
\label{subsec:Simulation_Details}

\begin{figure}[htb]
\centering
\includegraphics[scale=0.5]{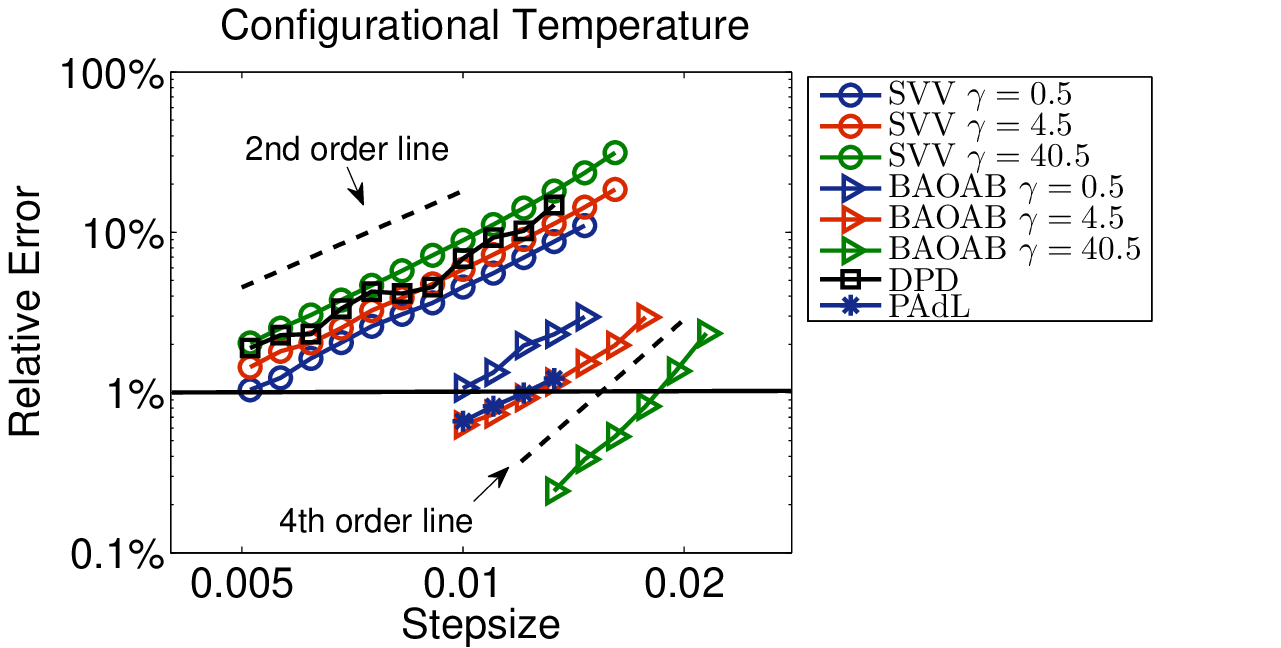}
\caption{\small (Color online) Double logarithmic plot of the relative error, i.e., the ratio between the absolute error of configurational temperature $f_T$ \eqref{eq:Config_Temp_f} and the preset temperature $T$, against stepsize by using various numerical methods introduced in Section~\ref{sec:Numerical_Methods} with a variety of friction coefficients in a standard setting of polymer melts as described in Section~\ref{subsec:Simulation_Details}. Note that the relative error of DPD and PAdL appears to show little dependence on the friction coefficients, thus the result of $\gamma=0.5$ only is shown. The system was simulated for 1000 reduced time units in each case but only the last 80\% of the snapshots were collected to calculate the static quantity $f_T$. Five different runs were averaged to further reduce the sampling errors. The stepsizes tested began at $h=0.005$ and were increased incrementally by 10\% until all methods became unstable. The horizontal solid black line indicates $1\%$ relative error in sample mean~\eqref{eq:sample_mean} accuracy of configurational temperature, based on which the stepsizes for each method were chosen in equilibrium simulations, unless otherwise stated. Dashed black lines represent the second and fourth order convergence to the invariant distribution. }
\label{fig:Melts_Comp_CT}
\end{figure}

A popular bead-spring model originally proposed by Kremer and Grest~\cite{Grest1986,Kremer1990} is used in our simulations. The system is composed of $M$ identical linear chains with $N$ beads each in a cubic box with periodic boundary conditions~\cite{Allen1989}. The total number of beads is $N_\mathrm{t}=MN$ in that case.
Excluded volume interactions between all $N_\mathrm{t}$ beads are included via a truncated Lennard-Jones potential:
\begin{equation*}\label{eq:Potential_LJ}
  U_{\mathrm{LJ}}(r_{ij})=
  \begin{cases}
    \displaystyle 4 \epsilon \left[ \left( \frac{\sigmaLJ}{r_{ij}} \right)^{12} \!\!\!-\! \left( \frac{\sigmaLJ}{r_{ij}} \right)^{6} \!-\! \left( \frac{\sigmaLJ}{r_{\mathrm{c}}} \right)^{12} \!\!+ \left( \frac{\sigmaLJ}{r_{\mathrm{c}}} \right)^{6} \right] \, , & r_{ij}<r_{\mathrm{c}} \, ;\\
    \quad \quad \quad \quad \quad \quad \quad \quad \ 0 \, , & r_{ij}\geq r_{\mathrm{c}} \, ,
  \end{cases}
\end{equation*}
where $r_{ij}=\|\q_{i}-\q_{j}\|$ denotes the distance between two beads $i$ and $j$, $\epsilon$ and $\sigmaLJ$ are two constants that set the energy and length scales of the beads, respectively, in reduced Lennard-Jones units, and $r_{\mathrm{c}}$ is the cutoff radius typically chosen as $r_{\mathrm{c}} = 2^{1/6}\sigmaLJ$ such that only the repulsive part of the potential is considered. This potential is also known as the Weeks--Chandler--Andersen potential~\cite{Weeks1971}.

Adjacent $N$ beads along the same polymer interact, in addition, via the finitely extensible nonlinear elastic (FENE) potential:
\begin{equation*}\label{eq:Potential_FENE}
  U_{\mathrm{FENE}}(r_{ij})=
  \begin{cases}
    \displaystyle - \frac{1}{2} k R^{2}_{\mathrm{max}} \ln \left[ 1 - \left( r_{ij}/R_{\mathrm{max}} \right)^{2} \right] \, , & r_{ij}<R_{\mathrm{max}} \, ;\\
    \quad \quad \quad \quad \quad \quad \ \infty \, , & r_{ij}\geq R_{\mathrm{max}} \, ,
  \end{cases}
\end{equation*}
where $k=30\epsilon/\sigmaLJ^{2}$ represents the spring coefficient and $R_{\mathrm{max}} = 1.5\sigmaLJ$ determines the maximum length of a bond. This choice of parameters ensures that chains do not cross each other and it allows for a reasonable large integration timestep~\cite{Kremer1990}. The system is thermostatted as described in Section~\ref{sec:Numerical_Methods}.

Overall, the total potential energy of the system is defined as
\begin{equation}
  U(\q) = \sum^{N_\mathrm{t}-1}_{i=1} \sum^{N_\mathrm{t}}_{j=i+1} \left[ U_{\mathrm{LJ}}(r_{ij}) + U_{\mathrm{FENE}}(r_{ij}) \right],
\end{equation}
and the total potential of a bond, $U_{\mathrm{LJ}}(r_{ij}) + U_{\mathrm{FENE}}(r_{ij})$ gives rise to a mean bond length $\langle r_{ij}\rangle\approx 0.97\,\sigmaLJ$ between adjacent beads $i$ and $j$ for $\epsilon=\kB T=$ 1.

A simple and popular choice of the weight function as in Ref.~\citen{Groot1997} was adopted in this article:
\begin{equation}\label{eq:Weight_Function_R}
  \omega^{\mathrm{R}}_{ij} = \omega^{\mathrm{R}}(r_{ij})=
  \begin{cases}
  \displaystyle 1-\frac{r_{ij}}{r_{\mathrm{c}}} \, , & r_{ij}<r_{\mathrm{c}} \, ;\\
  \quad 0 \, , & r_{ij}\geq r_{\mathrm{c}} \, .
  \end{cases}
\end{equation}

A system (bead number density $\rho_{\rm d}=0.84$) consisting of $M=30$ identical linear chains with $N=20$ beads (unit mass $m$) on each chain was simulated, where the following parameter set was used: $k_{\mathrm{B}}=\epsilon=\sigmaLJ=\mi =$ 1 (defining reduced units) at $T=1$ in reduced units. The thermal mass in PAdL was initially chosen as $\mu=10$. It should be noted that the simulated system is usually referred to as ``unentangled'' since $N \leq $ 85 \cite{Kroeger1993}. Pre-equilibrated initial configurations were obtained using an existing hybrid approach~\cite{Kroeger1999}: Initially, a mixture of phantom and excluded volume FENE chains were placed randomly into the simulation box at a density that exceeds the target density. A subsequent molecular dynamics algorithm with integration time step, force shape and force strength control was used to achieve a prescribed minimum distance (here 0.9) between all pairs of particles, while attempting to maintain local and global characteristics (such as the form factor) of the chain conformations. During this process the most inefficient chains were removed from the system, until the target density was reached. The initial momenta were independent and identically distributed (i.i.d.) normal random variables with mean zero and variance $k_{\mathrm{B}}T$. Unless otherwise stated, the system was simulated for 1000 reduced time units in each case but only the last 80\% of the snapshots were collected to calculate various quantities described in the preceding section.

\begin{figure}[tb]
\centering
\includegraphics[scale=0.3]{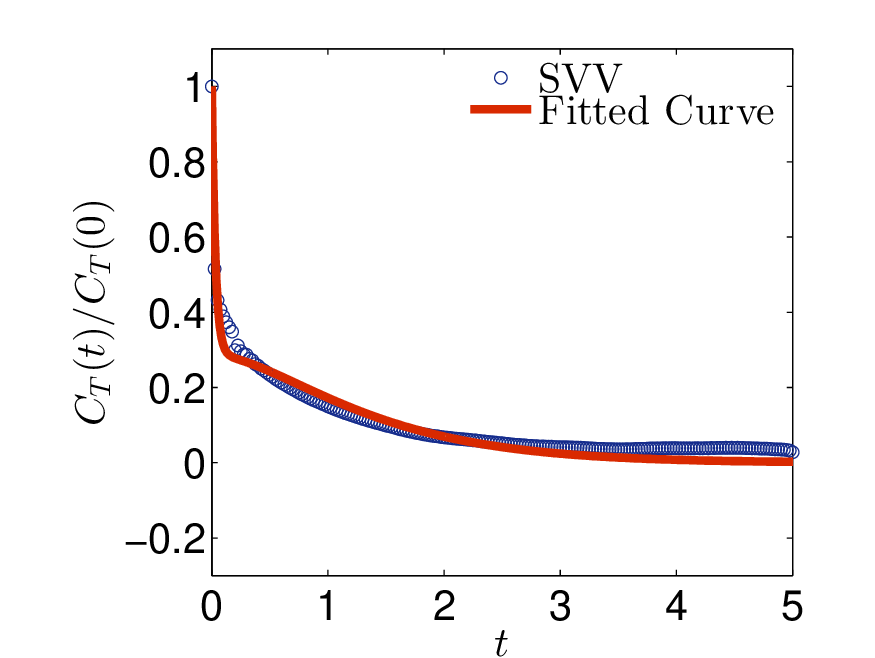}
\includegraphics[scale=0.3]{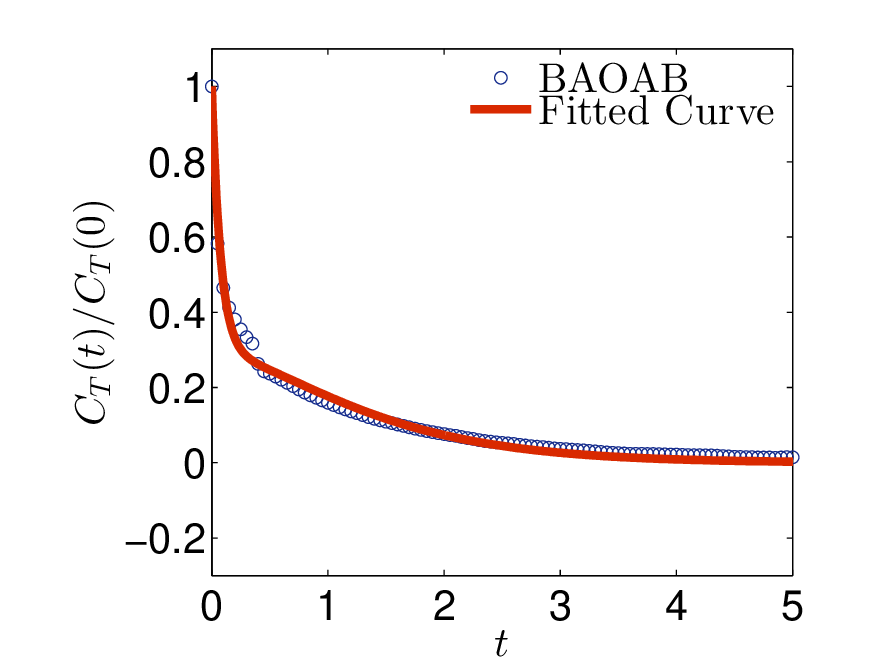}
\includegraphics[scale=0.3]{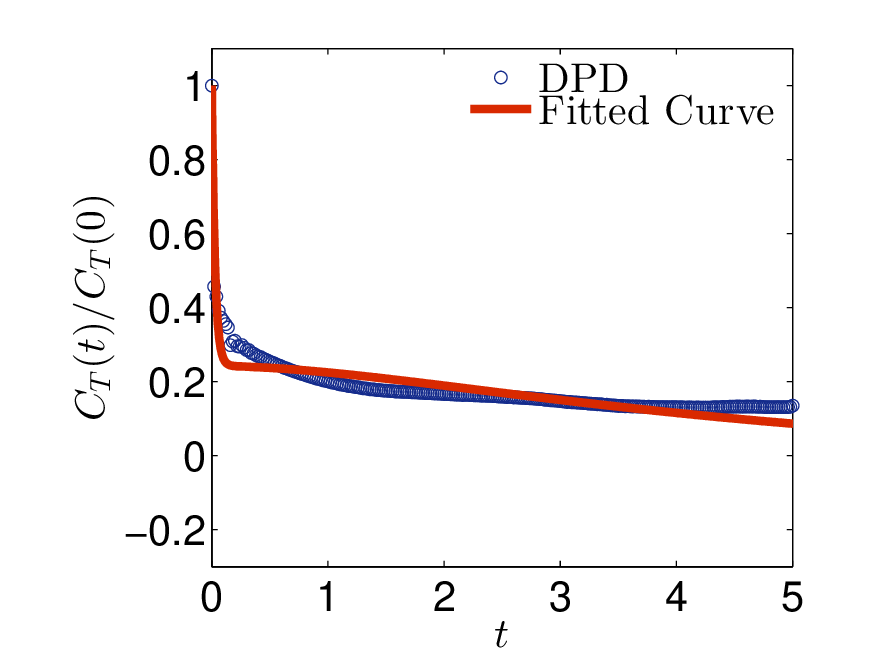}
\includegraphics[scale=0.3]{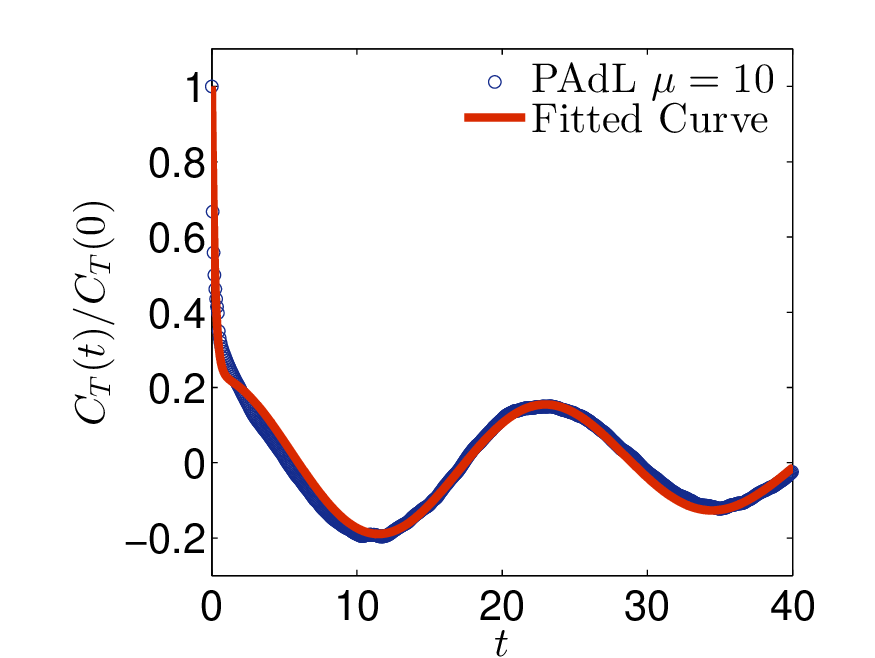}
\includegraphics[scale=0.3]{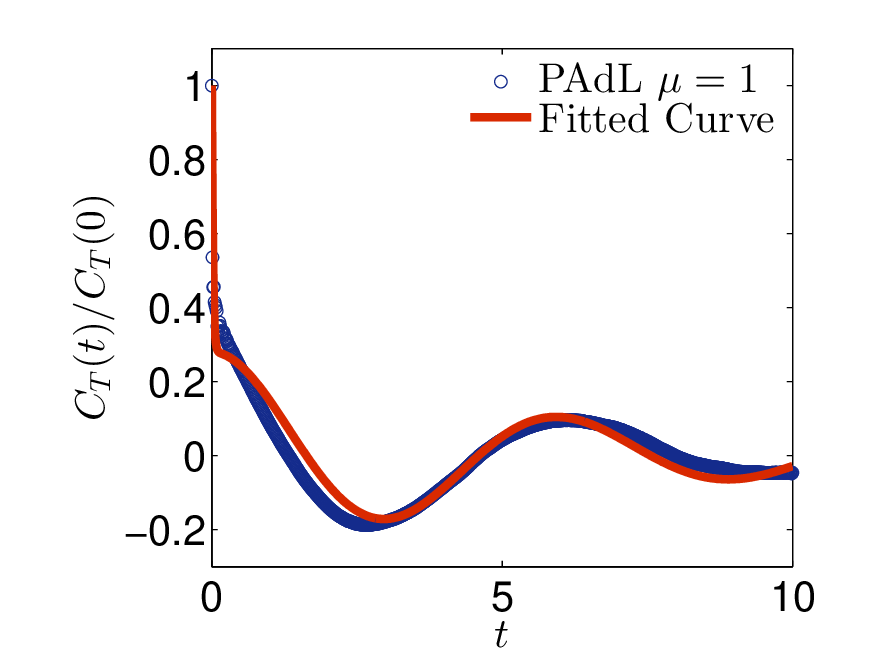}
\includegraphics[scale=0.3]{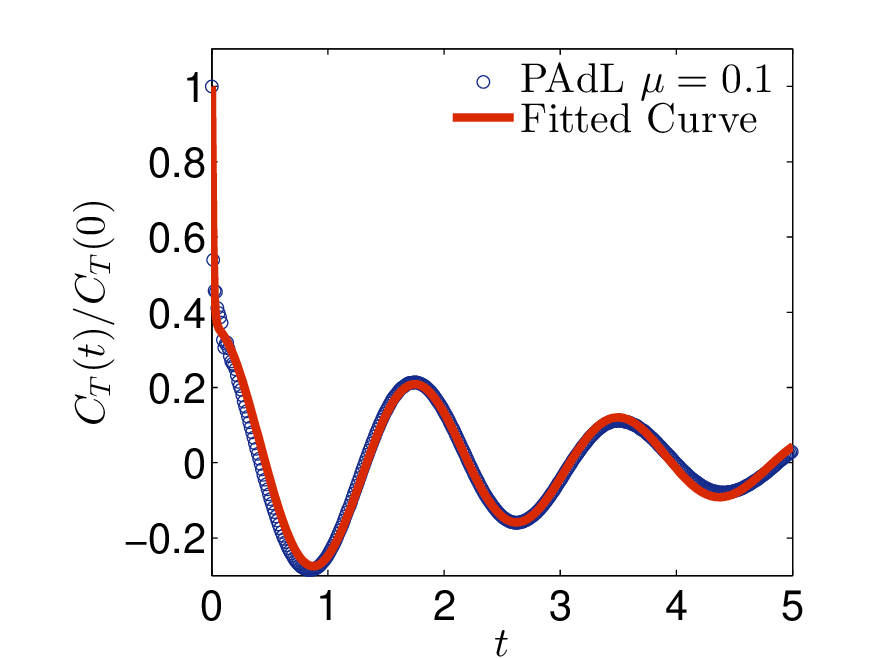}
\caption{\small (Color online) Weighted sum fittings, employing~\eqref{eq:Fitting_Weighted_Sum}, of the normalized configurational temperature autocorrelation function~\eqref{eq:AF} by using various methods with fixed stepsizes shown in Table~\ref{table:SAC_gamma_0d5_RE_1} in the low friction regime of $\gamma=0.5$ with sample mean accuracy of $\sim 1\%$ relative error in configurational temperature. Note that the horizontal axis is scaled to the reduced physical time, rather than the number of steps, for convenience. }
\label{fig:Melts_fittingCTAF_RE_1}
\end{figure}

\subsection{Equilibrium} \label{subsec:Equilibrium}

As a verification of our equilibrium simulations, we first investigate the structural quantities obtained by using the SVV, BAOAB, DPD, and PAdL methods introduced in Section~\ref{sec:Numerical_Methods}. Among all the methods, at a stepsize of $h=0.01$, we obtain an (time and chain) averaged squared end-to-end distance of $\langle R^{2}_{\mathrm{ee}} \rangle = 29.46$ and an averaged squared radius of gyration~\cite{Kremer1990} of $\langle R^{2}_{\mathrm{g}} \rangle = 4.87$ in the low friction regime of $\gamma = 0.5$,  in perfect agreement with the results of Kremer and Grest~\cite{Kremer1990}.

Figure~\ref{fig:Melts_Comp_CT} compares the configurational temperature~\eqref{eq:Config_Temp} control for a variety of methods with a range of friction coefficients. Note that SVV and BAOAB are two different splitting methods of Langevin dynamics. However, it can be clearly seen that, while maintaining a similar accuracy control of the configurational temperature, the BAOAB method allows the use of much larger (at least doubled) stepsizes compared to the SVV method, especially in the large friction limit ($\gamma=40.5$), where a superconvergence (i.e., a fourth order convergence, indicated by the dashed black line in the figure, to the invariant distribution) result was observed (as in Refs.~\citen{Leimkuhler2013,Leimkuhler2015a}).  All the other methods tested show second order convergence according to the dashed order line. This again illustrates the importance of optimal design of numerical methods. It is also interesting to note that the relative error slightly rises as we increase the friction coefficient for the SVV method whereas, in the BAOAB method, the relative error decreases.

\begin{table}[tb]
\small
  \caption{\ Comparisons of the sampling efficiency of various numerical methods quantified by the ``effective sample size'', $N_\mathrm{s}/\tau_{T}$, in the low friction regime of $\gamma=0.5$ with similar sample mean $\langle f_{T}\rangle$ accuracy of $\sim 1\%$ relative error in configurational temperature. For all entries the total simulation time was $N_s h \approx 800$, where $N_s$ and $h$ respectively represent the number of samples and the stepsize, and $I$ denotes the integrated normalized autocorrelation function~\eqref{eq:Fitting_Integral} of $f_T$. The DPD method was computed by using Shardlow's splitting method (i.e., the DPD-S1 scheme)~\cite{Shardlow2003}. The simulation details of the table are the same as in Figure~\ref{fig:Melts_Comp_CT}.}
  \label{table:SAC_gamma_0d5_RE_1}
  \begin{tabular*}{1.0\textwidth}{@{\extracolsep{\fill}}llrlrrr}
    \hline
    \textbf{Method} &
    \multicolumn{1}{c}{\textbf{$h$}} &
    \multicolumn{1}{c}{\textbf{$N_\mathrm{s}$}} &
    \multicolumn{1}{c}{\textbf{$\langle f_{T}\rangle$}} &
    \multicolumn{1}{c}{\textbf{$I$}} &
    \multicolumn{1}{c}{\textbf{$\tau_{T}$}} &
    \multicolumn{1}{c}{\textbf{$N_\mathrm{s}/\tau_{T}$}} \\
    \hline
    SVV   & 0.005 & 160001 & 1.0105 & 0.4412 & 175.5 & 911.7 \\
    BAOAB & 0.01  & 80001  & 1.0134 & 0.4863 & 96.3  & 830.7 \\
    DPD   & 0.004 & 200001 & 1.0093 & 1.1270 & 562.5 & 355.6 \\
    PAdL $\mu=10$ & 0.012  & 66668 & 0.9903 & 0.2703 & 44.1 & 1511.7 \\
    PAdL $\mu=1$  & 0.012  & 66668 & 0.9902 & 0.0959 & 15.0 & 4444.5 \\
    PAdL $\mu=0.1$ & 0.012 & 66668 & 0.9902 & 0.0249 & 3.2  & 20833.8 \\
    \hline
  \end{tabular*}
\end{table}

While the relative error of both SVV and BAOAB methods depends on the friction coefficients in Langevin dynamics,  the two pairwise thermostats (i.e., DPD and PAdL) appear to show little dependence on the friction coefficients. Although it seems that the DPD method is as accurate as SVV, the PAdL method is superior to both of them (even slightly better than BAOAB at low friction, i.e., $\gamma=0.5$).

The accuracy and rates of convergence for each method (and for each observable) depend in a nontrivial way on stepsize and so we cannot expect to use the same stepsize for different numerical integrators.
In performing comparisons, it is crucial to develop a careful procedure to quantify sampling convergence in relation to the accuracy desired.
In our studies we use a fixed accuracy threshold to select the stepsize for each method (in equilibrium) and then, for this choice of stepsize, which will be different for each method,  we use the configurational temperature SAC (see Section~\ref{subsubsec:SAC}) to estimate the convergence rate.  The detailed protocol is as follows:
\begin{enumerate}
  \item Choose a suitable observable (for instance, the configurational temperature throughout this article);
  \item Determine the stepsize, $h$, for each method by requiring an identical accuracy of the sample mean $\langle f_{T}\rangle$ (for instance, 1\% relative error, marked as the horizontal solid black line, as shown in Figure~\ref{fig:Melts_Comp_CT});
  \item The number of samples, $N_\mathrm{s}$, for each method is subsequently specified as the total simulation time is kept fixed;
  \item Approximate the SAC via Eq.~\eqref{eq:Fitting_Integral}, which is based on a weighted sum fitting of the normalized autocorrelation function.
  \item Calculate the effective sample size, $N_\mathrm{s}/\tau_{T}$, for each method, which characterize the sampling efficiency.
\end{enumerate}

\begin{figure}[tb]
\centering
\includegraphics[scale=0.5]{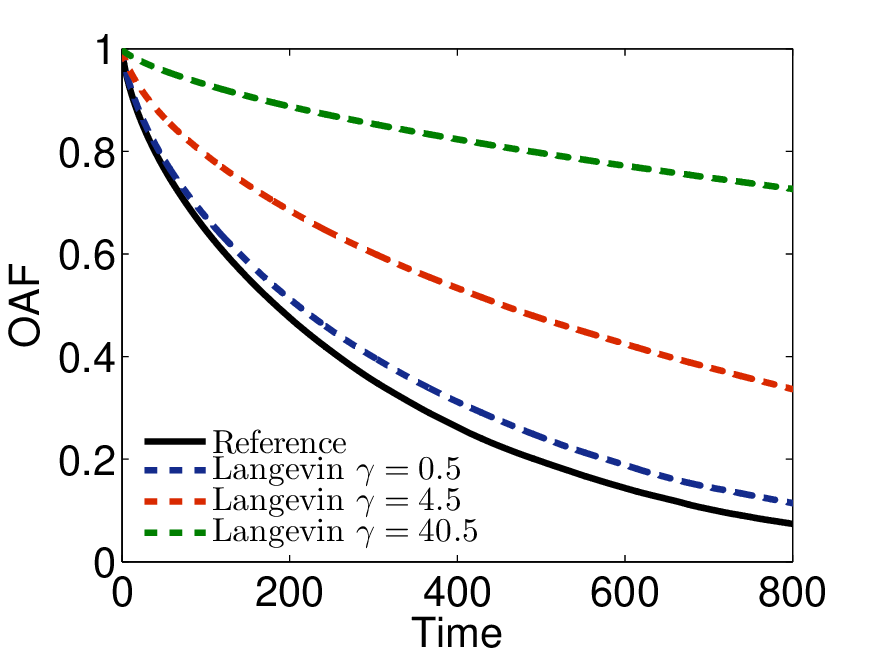}
\includegraphics[scale=0.5]{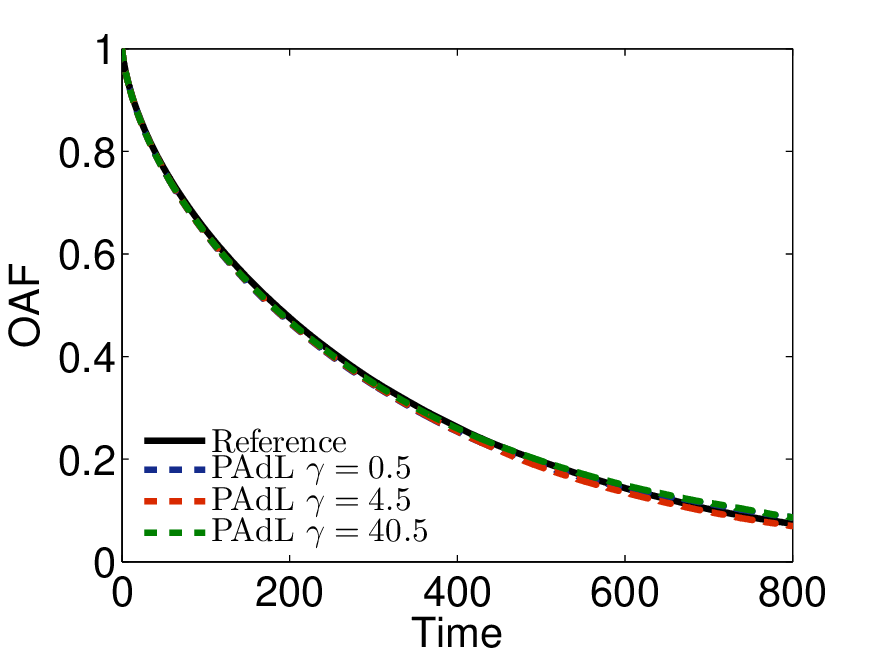}
\caption{\small (Color online) Comparisons of the normalized orientational autocorrelation function (OAF), $C_\mathrm{ee}(t)/C_\mathrm{ee}(0)$, of the end-to-end vector of polymer chains in a melt between Langevin dynamics (left) and the PAdL method (right) with three different values of the friction coefficient. Note that PAdL and DPD exhibit indistinguishable behavior and thus only the result of the former was shown. The same stepsize of $h=0.01$ was used for all methods as the focus here is to study the autocorrelation decays (in fact, reducing the stepsize leaves the autocorrelation decays indistinguishable). 100 different runs were averaged to reduce the sampling errors after the system was well equilibrated. The solid black line is the reference decay obtained by using Hamiltonian dynamics (i.e., switching off the thermostat, $\gamma=0$), which used exactly the same initial conformations and velocities as alternative stochastic dynamics.}
\label{fig:Melts_OAF}
\end{figure}

In order to measure the sampling efficiency of the various methods (three different values of the thermal mass, $\mu$, of PAdL are included), we plot the normalized configurational temperature autocorrelation function (CTAF) and its corresponding fitted curve based on the weighted sum~\eqref{eq:Fitting_Weighted_Sum} in Figure~\ref{fig:Melts_fittingCTAF_RE_1}. The stepsize, corresponding to the $\sim 1\%$ relative error case, for each method was determined with the help of the horizontal solid black line in Figure~\ref{fig:Melts_Comp_CT}. As can be seen from the normalized autocorrelation functions in Figure~\ref{fig:Melts_fittingCTAF_RE_1}, the samples of various methods decorrelate very differently. It is interesting to note that while both Langevin dynamics and DPD exhibit ``monotonic-like'' decays, PAdL (with a wide range of the thermal mass) oscillates, which we believe is due to the presence of the additional Nos\'{e}--Hoover control.

As can be seen from Table~\ref{table:SAC_gamma_0d5_RE_1}, the SAC ($\tau_{T}$) of DPD is significantly larger than alternatives. Moreover, the SAC of BAOAB is about half of that of SVV, which is due to the fact that BAOAB can use about double the stepsize of SVV and thus only about half as many samples are needed to achieve a similar sample mean accuracy of the configurational temperature. Although the SAC of PAdL with $\mu=10$ is already smaller than either Langevin dynamics or DPD, the SAC of PAdL could be decreased further by lowering the value of the thermal mass $\mu$. This should not come as a great surprise since $\mu$ determines how strongly the negative feedback loops~\eqref{eq:PNHL_G} couple with the physical system. Therefore, the PAdL method has the ability of controlling the sampling efficiency while others not.

\begin{figure}[tb]
\centering
\includegraphics[scale=0.5]{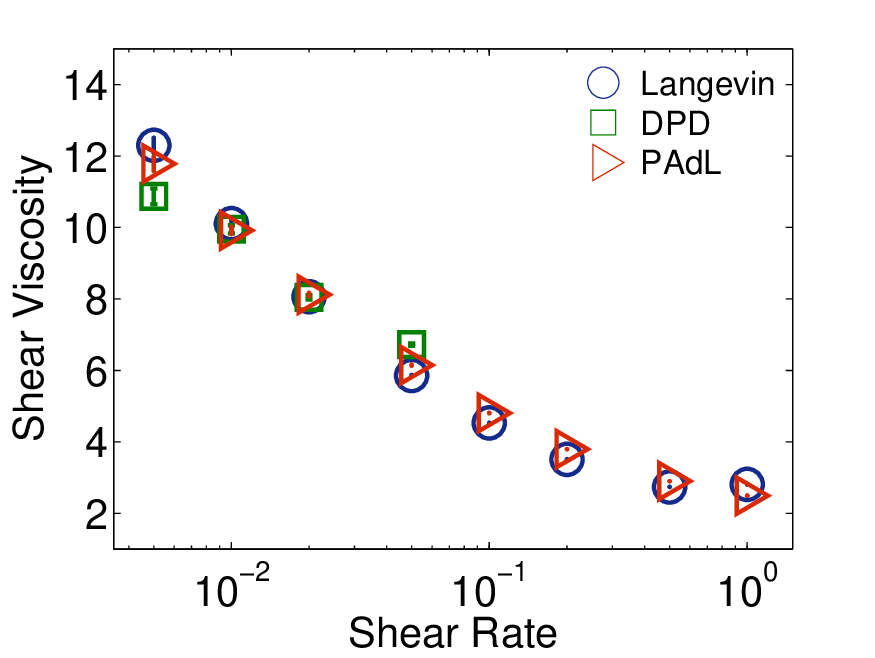}
\caption{\small (Color online) Comparisons of the computed shear viscosity in a standard setting of polymer melts as described in Section~\ref{subsec:Simulation_Details} against shear rate by using various methods (in the low friction regime of $\gamma=0.5$) with Lees--Edwards boundary conditions. The simulation details of the figure are the same as in Figure~\ref{fig:Melts_Comp_CT} except the same stepsize of $h=0.01$ was used for all methods. Note that the error bars were included but could be seen with relatively small shear rates only. Note also that we did short runs to highlight the deviations, while the errors decrease further and the viscosity reaches the Newtonian plateau at small shear rates upon increasing the length of the runs, as is well known from previous studies.}
\label{fig:Melts_Shear_Viscosity_Rate}
\end{figure}

Since different stepsizes, which result in different sample counts ($N_\mathrm{s}$) when the total simulation time is fixed, were used for different methods in order to achieve a similar sample mean accuracy, one should further compare the sampling efficiency by  computing the ``effective sample size'' ($N_\mathrm{s}/\tau_{T}$) instead of just the SAC ($\tau_{T}$), which corresponds to cases where $N_\mathrm{s}$ is identical in each method. One can see from Table~\ref{table:SAC_gamma_0d5_RE_1} that the effective sample sizes of SVV and BAOAB are very close to each other since they are just two different splitting methods of the same stochastic (Langevin) dynamics. While the effective sample size of DPD is roughly half that of Langevin dynamics (either SVV or BAOAB), PAdL with $\mu=10$ is over 50\% larger than the latter. Further reducing the value of the thermal mass in PAdL leads to an even more efficient sampling than alternatives: the effective sample size of PAdL with $\mu=1$ is more than four times that of DPD and Langevin dynamics; with $\mu=0.1$, this increases to a factor of 20.    There is a limit to how much $\mu$ can be reduced, however, without introducing numerical instability and thus requiring a smaller timestep. Note that although the thermal mass $\mu$ in PAdL has a strong influence on the sampling efficiency, it appears that the sampling accuracy depends little on it (for instance, the long term behavior of PAdL with a wide range of the thermal mass $\mu$ is almost indistinguishable in Figure~\ref{fig:Melts_Comp_CT}). Therefore, unless otherwise stated, $\mu=0.1$ will be used in subsequent comparisons.

The characterization of the relaxation of polymer chains in a melt corresponding to different dynamics was compared and plotted in Figure~\ref{fig:Melts_OAF}. In particular, we measured the orientational autocorrelation function (OAF) of the end-to-end vector of polymer chains defined in Section~\ref{subsubsec:Correlations}. It is believed that,  for such a dense system, the long-time diffusion depends only on the interactions between beads and does not arise from the associated thermostat~\cite{Kremer1990}. Therefore, the reference decay was calculated by using Hamiltonian dynamics (i.e., switching off the thermostat, $\gamma=0$). Since the OAFs obtained by SVV and BAOAB are almost indistinguishable, the SVV method was used for Langevin dynamics. (Note that in what follows, unless otherwise stated, Langevin dynamics was calculated by using the benchmark SVV method).

It can be seen from Figure~\ref{fig:Melts_OAF} (left) that the OAF of Langevin dynamics depends strongly on the friction coefficient. To be more precise, the OAF starts to (significantly) deviate from the reference decay as we increase the friction coefficient. Although a relative small friction (i.e., $\gamma=0.5$ in this parameter setting) was suggested in Ref.~\citen{Kremer1990} to not only minimize the effects of the Langevin thermostat but also to be large enough to stabilize the system in the long time limit, visible discrepancies were still observed in the case of $\gamma=0.5$ in our numerical experiments. In stark contrast, the OAFs of the PAdL method in a wide range of the friction coefficients are almost indistinguishable from the reference decay as shown in Figure~\ref{fig:Melts_OAF} (right). Very similar behavior was also observed in the DPD method, which implies that the projection of the interactions of both the dissipative and random forces (i.e., the thermostat) on to the line of centers (and thus the conservation of the momentum) may have played a role in preserving the correct relaxation behavior in the case of the pairwise thermostats. Since a relatively small friction has been widely used in the literature of polymer melts, in what follows we restrict our attention to comparing various methods in the low friction regime of $\gamma=0.5$.

\subsection{Nonequilibrium}

The stepsize for each method was chosen according to certain sample mean accuracy threshold (e.g., $\approx 1\%$ relative error in configurational temperature) when examining the sampling efficiency in the previous subsection. However, we are more interested in investigating the stability issues in nonequilibrium simulations. Thus in what follows we fix the stepsize of $h=0.01$, which is close to the stability threshold, for all methods.

\begin{figure}[tb]
\centering
\includegraphics[scale=0.5]{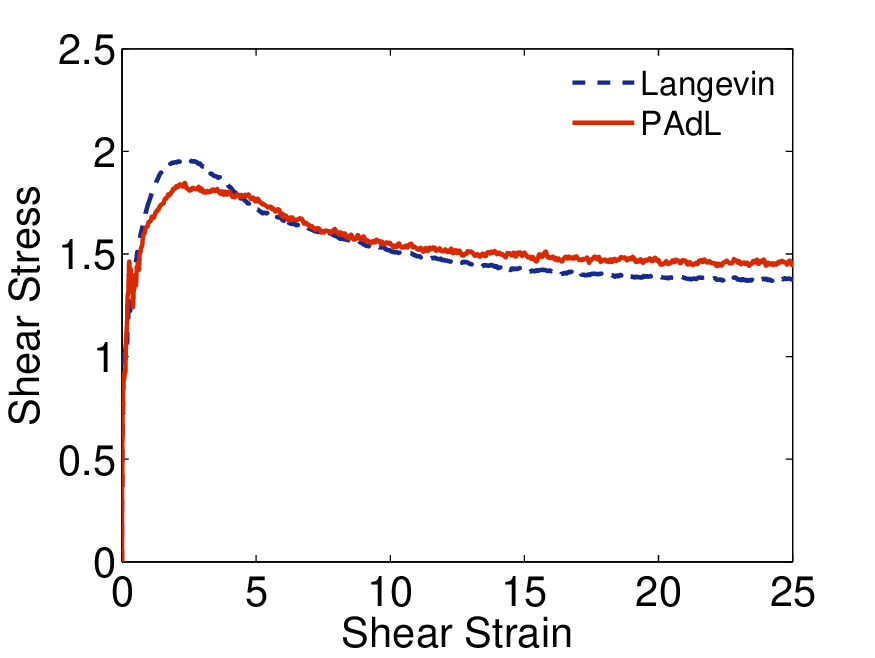}
\caption{\small (Color online) Plot of the relations of shear stress and shear strain, which is the product of shear rate and simulation time, between Langevin dynamics and PAdL (in the low friction regime of $\gamma=0.5$) for the polymer melts modeled using Lees--Edwards boundary conditions with a shear rate of $\dot{\gamma}=0.5$. The same stepsize of $h=0.01$ was used for both methods. 10000 different runs were averaged to obtain relatively smooth curves.}
\label{fig:Melts_Shear_Stress_Strain}
\end{figure}

\begin{figure}[tb]
\centering
\includegraphics[scale=0.5]{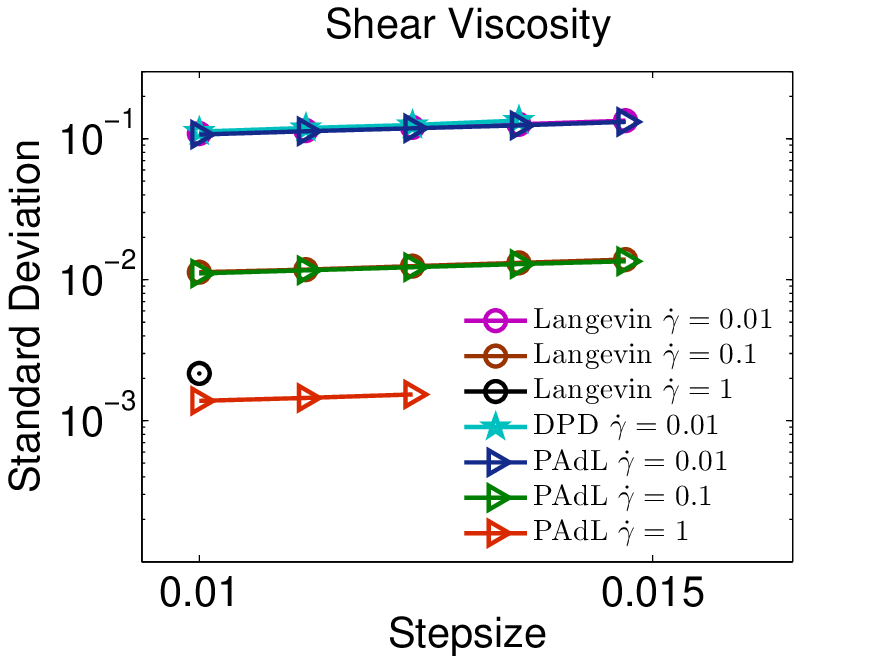}
\caption{\small (Color online) Double logarithmic plot of the standard deviation in the computed shear viscosity at the specified rates against stepsize by using various methods (in the low friction regime of $\gamma=0.5$) in the presence of Lees--Edwards boundary conditions. The simulation details of the figure are the same as in Figure~\ref{fig:Melts_Comp_CT} except the stepsizes tested began at $h=0.01$.}
\label{fig:Melts_Shear_Viscosity_SD}
\end{figure}

The shear viscosity was extracted by using the formula~\eqref{eq:Shear_Viscosity} outlined in Section~\ref{subsubsec:Shear_Viscosity} and plotted in Figure~\ref{fig:Melts_Shear_Viscosity_Rate}. All methods appear to show similar behavior except that the range of shear rates in DPD is greatly limited (that is, the largest usable shear rate in DPD is $\dot{\gamma}=0.08$, compared with $\dot{\gamma}=1.2$ in both Langevin and PAdL). The error bars were indeed included, however they were relatively very small (particularly in the large shear rate regime) and thus not visible.

\begin{figure}[tb]
\centering
\includegraphics[scale=0.5]{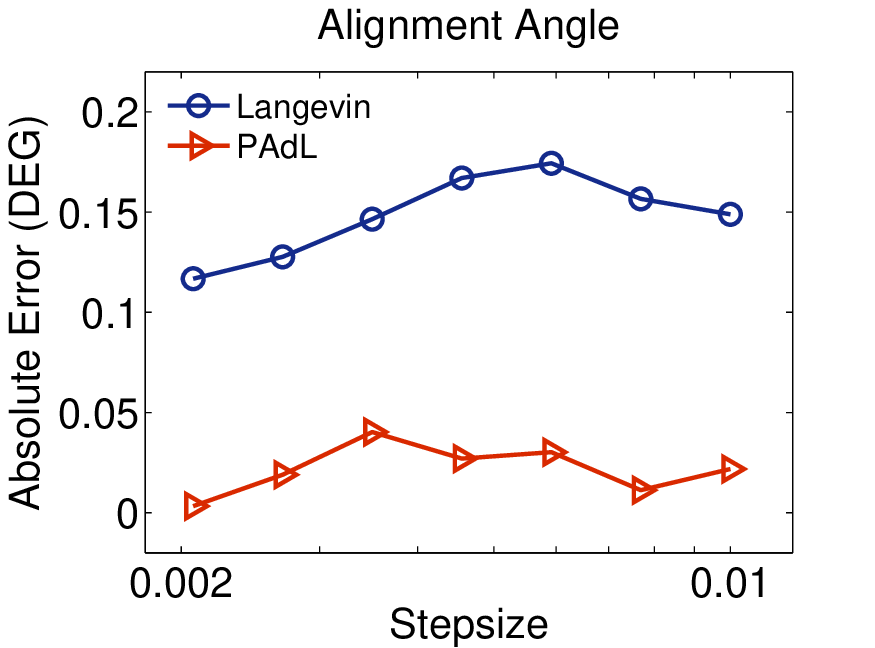}
\caption{\small (Color online) For the polymer melts modeled using Lees--Edwards boundary conditions, the absolute error (in degrees) of the flow alignment angle~\eqref{eq:Alignment_Angle} produced by the PAdL and Langevin algorithms (in the low friction regime of $\gamma=0.5$) is plotted against the stepsize (in semi-log scale) for a shear rate of $\dot{\gamma}=1$. The reference value of the alignment angle was obtained by using the PAdL method with a sufficiently small stepsize of $h=0.001$ (20 different runs). The simulation details of the figure are the same as in Figure~\ref{fig:Melts_Comp_CT} except 20 different runs were averaged to further reduce the sampling errors.}
\label{fig:Melts_Shear_1_AA}
\end{figure}

In Figure~\ref{fig:Melts_Shear_Stress_Strain}, we plot the evolution of the shear stress against the shear strain for PAdL and Langevin. While PAdL was perfectly stable for all runs performed, Langevin dynamics occasionally becomes unstable with a shear rate of $\dot{\gamma}=1$, therefore, a smaller shear rate of $\dot{\gamma}=0.5$ was used here instead in order to provide comparisons. As can be seen from the figure, for both methods, as the shear strain increases, the shear stress rises rapidly from zero to its maximum at shear strain of around 2.5 before relaxing to its steady state, consistent with the results in Figure~\ref{fig:Melts_Shear_Viscosity_Rate}. At startup the system tends to transform affinely and builds up stress before relaxation takes over. The faster the shear rate, the more likely the shear stress maximum occurs, as affine shear deformation.  During this phase, particles are forced into close proximity resulting in strong Lennard-Jones repulsion and quickly producing enormous shear stress. However, the maximum is not expected, and was indeed not observed by us, at small rates. Overall, the behavior is in good agreement with previous studies~\cite{Cao2015,Hoy2007,Wagner1980,Bird1987}. We also investigated the evolution (including DPD whenever possible) with a wide range of shear rates and did not observe significant differences between the methods.

Figure~\ref{fig:Melts_Shear_Viscosity_SD} plots the standard deviation in the computed shear viscosity over a wide range of shear rates by using various methods subject to Lees--Edwards boundary conditions. Once again, all methods behave very similarly. As we increase the shear rate, the standard deviation decreases, which is consistent with the observations in Figure~\ref{fig:Melts_Shear_Viscosity_Rate}. Note that the DPD method appears only in the smallest shear rate case. In the regimes of small and moderate shear rates, the standard deviations of all/both methods are very similar. However, in the high shear rate ($\dot{\gamma}=1$) case, Langevin dynamics has not only a visibly ($\approx $ 60\%) larger standard deviation but also a smaller range of stepsizes usable than the PAdL method.

We further investigate in Figure~\ref{fig:Melts_Shear_1_AA} the stepsize effects on the alignment angle of the polymer chains subject to LEBC at a relatively high shear rate of $\dot{\gamma}=1$. It appears that while the absolute error of the alignment angle of PAdL  remains below around 0.04 degrees, the corresponding error of Langevin dynamics is around six times larger. This clearly demonstrates the superiority of PAdL over Langevin dynamics in nonequilibrium simulations especially at relatively high shear rates.

\section{Conclusions}
\label{sec:Conclusions}

We have reviewed a variety of numerical methods (SVV and BAOAB of Langevin dynamics, DPD, and PAdL) that can be used to simulate polymeric systems. We have systematically compared those methods in terms of accuracy, efficiency, and stability both in equilibrium and nonequilibrium settings.

In terms of sampling accuracy in equilibrium simulations, we have observed that the BAOAB and PAdL methods outperform the SVV and DPD methods in a wide range of friction coefficients. We have also discovered that while perfectly matching the reference decay, the OAF, which characterizes the orientational relaxation of the polymer chains, of both pairwise (momentum-conserving) thermostats (DPD and PAdL) has little dependence on the friction coefficient in a wide range. On the other hand, the OAF of Langevin dynamics strongly depends on the friction coefficient, moreover, a clear discrepancy was observed even for the commonly used low friction of $\gamma=0.5$. We have further developed a careful procedure to quantify the sampling efficiency of various methods. By comparing the effective sample size, we found that PAdL substantially outperforms alternatives, particularly with a relatively small thermal mass of $\mu=0.1$, for which remarkably about twenty times increase in the effective sample size was achieved in comparison to alternative approaches (see Table~\ref{table:SAC_gamma_0d5_RE_1}).

We are more focused on investigating the stability issues in nonequilibrium simulations. We have demonstrated that, with a stepsize of $h=0.01$, the largest usable shear rate was around $\dot{\gamma}=0.08$ for DPD, compared with $\dot{\gamma}=1.2$ for both Langevin and PAdL, in a standard setting of polymer melts as described in Section~\ref{subsec:Simulation_Details}. Thus, in agreement with previous studies~\cite{Fedosov2010}, DPD is not recommended for nonequilibrium simulations, when the mean flow dissipation rates begin to overwhelm the thermostat, limiting its use in practice for relatively large shear rates.    Between Langevin dynamics and PAdL, we have found that they perform rather similarly with relatively low shear rates. For the investigation of even smaller rates, where the FENE polymer exhibits a Newtonian plateau in the shear viscosity,  thermodynamically guided methods are more suitable~\cite{Ilg2009,Ilg2011}. Nevertheless, we have illustrated that while both methods share a similar relation of shear stress and shear strain, Langevin \mbox{dynamics} performed unreliably with a relatively high shear rate of $\dot{\gamma}=1$ at a stepsize of at least $h=0.01$---it not only produced a larger (by about 60\%) standard deviation of the computed shear viscosity than PAdL, but also resulted in a significantly larger (about six times) absolute error of the flow alignment angle.

\section*{Acknowledgements}

The authors thank Michael Allen, Gabriel Stoltz, Karl Travis, and anonymous referees for valuable suggestions and comments. XS and BL acknowledge the support of the Engineering and Physical Sciences Research Council (UK) through Grant No.\ EP/P006175/1. Part of this work was done during XS and BL's stay at the Institut Henri Poincar\'{e} - Centre \'{E}mile Borel during the trimester ``Stochastic Dynamics Out of Equilibrium''. XS and BL thank this institution for hospitality and support.



\begin{appendices}
  \renewcommand\thetable{\thesection\arabic{table}}
  \renewcommand\thefigure{\thesection\arabic{figure}}

\section{Integration schemes}
\label{sec:Appendix_Schemes}

We list here detailed integration steps for both DPD-S1 and PAdL methods described in the article. Verlet neighbor lists~\cite{Verlet1967} are used throughout each method in order to reduce the computational cost.

\subsection*{Shardlow's splitting method: DPD-S1}

The summations go over all interacting pairs within cutoff radius ($r_{ij}<r_{\mathrm{c}}$),
\begin{align*}
  \mathbf{p}_{i}^{n+1/4} & = \mathbf{p}_{i}^{n} - \sum_{j>i} H_{ij}(\mathbf{e}_{ij}^{n}\cdot \mathbf{v}_{ij}^{n})\mathbf{e}_{ij}^{n} + \sum_{j>i} {\bf J}_{ij}\, , \\
  \mathbf{p}_{j}^{n+1/4} & = \mathbf{p}_{j}^{n} + \sum_{j>i} H_{ij}(\mathbf{e}_{ij}^{n}\cdot \mathbf{v}_{ij}^{n})\mathbf{e}_{ij}^{n} - \sum_{j>i} {\bf J}_{ij}\, , \\
  \mathbf{p}_{i}^{n+2/4} & = \mathbf{p}_{i}^{n+1/4} + \sum_{j>i} {\bf J}_{ij} - \sum_{j>i} \frac{H_{ij}}{1+2H_{ij}} \left[(\mathbf{e}_{ij}^{n}\cdot \mathbf{v}_{ij}^{n+1/4})\mathbf{e}_{ij}^{n} + 2{\bf J}_{ij}\right], \\
  \mathbf{p}_{j}^{n+2/4} & = \mathbf{p}_{j}^{n+1/4} - \sum_{j>i} {\bf J}_{ij} + \sum_{j>i} \frac{H_{ij}}{1+2H_{ij}} \left[(\mathbf{e}_{ij}^{n}\cdot \mathbf{v}_{ij}^{n+1/4})\mathbf{e}_{ij}^{n} + 2{\bf J}_{ij}\right],
\end{align*}
where $H_{ij}\equiv \gamma \omega^{\mathrm{D}}(r_{ij}^{n}) h/2 $ and ${\bf J}_{ij}\equiv \sigma \omega^{\mathrm{R}}(r_{ij}^{n})\mathbf{e}_{ij}^{n} \sqrt{h}\mathrm{R}_{ij}^{n}/2$ with $\mathrm{R}_{ij}^{n}$ being normally distributed variables with zero mean and unit variance. 

\noindent For each particle $i$,
\begin{align*}
  \mathbf{p}_{i}^{n+3/4} & = \mathbf{p}_{i}^{n+2/4} + h\mathbf{F}_{i}^{\mathrm{C}}(\mathbf{q}^{n})/2 \, , \\
  \mathbf{q}_{i}^{n+1} & = \mathbf{q}_{i}^{n} + h\mathbf{v}_{i}^{n+3/4} \, , \\
  \mathbf{p}_{i}^{n+1} & = \mathbf{p}_{i}^{n+3/4} + h\mathbf{F}_{i}^{\mathrm{C}}(\mathbf{q}^{n+1})/2 \, ,
\end{align*}
where $\mathbf{F}^{\mathrm{C}}_{i}(\mathbf{q}) = -\nabla_{\mathbf{q}_{i}}U(\mathbf{q})$ are the total conservative forces acting on particle $i$ with configuration $\mathbf{q}$.

\subsection*{Pairwise adaptive Langevin thermostat: PAdL}

For each particle $i$,
\begin{align*}
  \mathbf{q}_{i}^{n+1/2} &=  \mathbf{q}_{i}^{n} + h\mathbf{v}_{i}^{n}/2 \, , \\
  \mathbf{p}_{i}^{n+1/4} &= \mathbf{p}_{i}^{n} + h\mathbf{F}_{i}^{\mathrm{C}}(\mathbf{q}^{n+1/2})/2 \, .
\end{align*}
\noindent The summations go over all interacting pairs within cutoff radius ($r_{ij}<r_{\mathrm{c}}$),
\begin{align*}
  \mathbf{p}^{n+2/4}_{i} &= \mathbf{p}^{n+1/4}_{i} + \frac{1}{2}\sum_{j>i} m\Delta v_{ij}(\mathbf{q}^{n+1/2},\mathbf{p}^{n+1/4},\xi^{n}) \mathbf{e}^{n+1/2}_{ij} \, , \\
  \mathbf{p}^{n+2/4}_{j} &= \mathbf{p}^{n+1/4}_{j} - \frac{1}{2}\sum_{j>i} m\Delta v_{ij}(\mathbf{q}^{n+1/2},\mathbf{p}^{n+1/4},\xi^{n}) \mathbf{e}^{n+1/2}_{ij} \, ,
\end{align*}
with
\begin{equation*}
  \begin{aligned}
  &\Delta v_{ij} =
   \begin{cases}
   \left( \mathbf{e}_{ij} \cdot \mathbf{v}_{ij} \right)\left[ e^{-\tilde{\tau} h/2} - 1 \right] + \sigma \displaystyle\sqrt{ \left[1-e^{-\tilde{\tau} h}\right]/\left(\xi m\right)} \,\mathrm{R}_{ij} \, , & \xi \neq 0 \, ;\\
  \left(2\sigma/m\right) \omega^{\mathrm{R}}(r_{ij}) \sqrt{h/2}\, \mathrm{R}_{ij} \, , & \xi = 0 \, ,
  \end{cases}
  \end{aligned}
\end{equation*}
where $\tilde{\tau}=2\xi \omega^{\mathrm{D}}(r_{ij}) / m$ and $\mathrm{R}_{ij}$ are normally distributed variables with zero mean and unit variance. 

\noindent For the additional variable $\xi$,
\begin{equation*}
  \xi^{n+1} =  \xi^{n} + hG(\mathbf{q}^{n+1/2},\mathbf{p}^{n+2/4})/2 \, ,
\end{equation*}
where
\begin{equation*}
    G(\mathbf{q},\mathbf{p}) = {\mu}^{-1}\sum_{i}\sum_{j>i}\omega^{\mathrm{D}}(r_{ij}) \left[ \left( \mathbf{v}_{ij}\cdot\mathbf{e}_{ij} \right)^{2} - 2k_{\mathrm{B}}T/m  \right] \, .
\end{equation*}

\noindent The following summations go over all interacting pairs within cutoff radius ($r_{ij}<r_{\mathrm{c}}$),
\begin{align*}
  \mathbf{p}^{n+3/4}_{i} &= \mathbf{p}^{n+2/4}_{i} + \frac{1}{2}\sum_{j>i} m\Delta v_{ij}(\mathbf{q}^{n+1/2},\mathbf{p}^{n+2/4},\xi^{n+1}) \mathbf{e}^{n+1/2}_{ij} \, , \\
  \mathbf{p}^{n+3/4}_{j} &= \mathbf{p}^{n+2/4}_{j} - \frac{1}{2}\sum_{j>i}  m\Delta v_{ij}(\mathbf{q}^{n+1/2},\mathbf{p}^{n+2/4},\xi^{n+1}) \mathbf{e}^{n+1/2}_{ij} \, .
\end{align*}

\noindent For each particle $i$,
\begin{align*}
  \mathbf{p}_{i}^{n+1} &= \mathbf{p}_{i}^{n+3/4} + h\mathbf{F}_{i}^{\mathrm{C}}(\mathbf{q}^{n+1/2})/2 \, , \\
  \mathbf{q}_{i}^{n+1} &=  \mathbf{q}_{i}^{n+1/2} + h\mathbf{v}_{i}^{n+1}/2 \, .
\end{align*}
An extension of the PAdL algorithm for systems with beads of different masses and identical friction coefficients can be found in
Ref.~\citen{Leimkuhler2016a}.

\end{appendices}

\bibliographystyle{is-abbrv}

\bibliography{refs}

\end{document}